\documentclass[10pt]{iopart}

\usepackage{iopams}  
\usepackage{graphicx}

\begin{document}

\title{Stochastic Hamiltonians for correlated electron models}

\author{Frederick Green}
%
\address
{School of Physics, The University of New South Wales,
Sydney, NSW 2052, Australia.}

 
\begin{abstract}
Microscopically conserving reduced models of many-body systems have a
long, highly successful history. Established perturbative theories
of this type are the random-phase approximation for  Coulomb fluids
and the particle-particle ladder model for nuclear matter.
There are also more physically comprehensive diagrammatic approximations,
such as the induced-interaction and parquet models. Notwithstanding
their explanatory power, some theories have lacked an explicit
Hamiltonian from which all significant system properties,
static and dynamic, emerge canonically. This absence can complicate
evaluation of the conserving sum rules, essential consistency checks
on the validity of any theory. In a series of papers Kraichnan introduced
a stochastic embedding procedure to generate explicit Hamiltonians
for common approximations for the full many-body problem. Existence of a
Hamiltonian greatly eases the task of securing fundamental identities in
such studies. I revisit Kraichnan's method to apply it to correlation
theories for which such a canonical framework has been missing.
I exhibit Hamiltonians for more elaborate correlated models incorporating
both long-range screening and short-range scattering phenomena. These
are relevant to the study of strongly interacting electrons and condensed
quantum systems broadly.
\end{abstract}


\section{Introduction}

In the last century, within a remarkably brief span, the study of
strongly correlated quantum systems witnessed a series of crucial
innovations. The earliest example, for the electron fluid, is the
random-phase approximation (RPA) of Bohm and Pines
\cite{DP1,DP2}, still a paradigm of many-body analysis today.
Within the perturbative, or diagrammatic, philosophy the RPA was
rapidly followed by formal developments from Martin and Schwinger
\cite{MS}, influencing the Green-function approach of Kadanoff and Baym
\cite{KB,GB}. Russian studies contributed in a major way
\cite{AGD}, the Keldysh formalism
\cite{keld,langr} being the most familiar and in widespread use.
Of the plentiful and thorough reference works surveying this vast area,
we cite four standard texts by Nozi\`eres and Pines
\cite{DP2}, Nozi\`eres
\cite{nozi}, Rickayzen
\cite{rick}, Mahan
\cite{mahan} and a more recent treatment by Coleman
\cite{PC}. These provide a valuable cross-section of different
perspectives and analytic techniques. 

The high-order perturbative models developed
in this period, with their more
specialized variants (as for superconductivity and superfluidity),
offered tractable approximations beyond
the long-ranged RPA to cover finer-scale, short-range correlations in
condensed systems from the electron gas, to nuclear matter, to the helium
fluids. The theories here in discussion are almost always cast in the
language of Green functions and their dynamical equations.

Despite their technical ingenuity and effectiveness, many diagrammatic
theories have had to be constructed bottom-up. A central conceptual
tool has been missing by way of an explicit Hamiltonian underpinning.
At times this has caused confusion around the interpretation of their
dynamical sum rules (essential tests of the conservation laws), not
to mention a level of {\em ad hoc} patchwork to try to fix these.

Amid these historical developments a canonical, top-down
strategy for building model Hamiltonians was devised by Kraichnan
\cite{RHK1,RHK2}, who turned his construction to the dominant
correlation theories of the time. Kraichnan's stochastic approach
to microscopic many-body dynamics revolutionized the different
field of turbulence theory
\cite{eyink} although,
while freely acknowledged to be of fundamental importance
to correlated quantum systems
\cite{GB}, his innovation and its potential do not appear
to have gained wide currency in the community.
To this writer's mind it is an opportunity to be fully grasped;
his sense is informed by an early and
time-consuming effort to prove a higher frequency-moment sum rule for a
correlated model of the electron gas
\cite{GNS3}. It should really be enough to demonstrate
the result once and for all and rely on the universality of the
procedure. Knowing the Hamiltonian particular to an approximation
would doubtlessly help.

Nondiagrammatic analyses have developed
side by side with diagram-based ones and, as with the latter,
they naturally start from
a fundamental originating Hamiltonian. Two of the
better-known nonperturbative approaches
are density-functional theory
\cite{dft1,dft2} and the coupled-cluster formalism
\cite{cc1,cc2};
but there, equally, a reduced Hamiltonian tailored to
some model may not emerge on the path to a tractable approximation.
Depending on which questions call for answers,
as with diagrammatic theories,
the lack of such a tool may have its disadvantages.

While the current work is diagrammatically oriented,
the model-Hamiltonian philosophy it adopts is wider than any
specialized approach, with possible implications for
nondiagrammatic approaches also.
As an illustration one recalls
the valuable insight of Jackson, Lande, and Smith
\cite{JLS} into the correspondence between
a specific, nonperturbative variational model and
a self-consistent diagrammatic one.
In such a context the possibility of constructing
a model Hamiltonian for one approximate description
would immediately reflect upon the other.

In this paper I return to Kraichnan's methodology to show how
it can be adapted readily to many-body formulations
beyond the approximations analyzed by him, and for which an
explicit Hamiltonian has not been available.
The next Section reviews the general method for building model
Hamiltonians. It should be stressed that, while the focus will be
on the uniform electron gas,
Kraichnan designed the method to apply equally well to any system,
finite or extended, with pair interactions.
To establish familiarity with the approach, Sec. III revisits the
classic approximations originally analyzed by him:
the random-phase model, Hartree-Fock, the ring approximation
and the particle-particle (Brueckner) ladder summation.
Sec. IV introduces more complex operations
for Hamiltonian models, both to single out RPA-related effects and
long-range screening and then to unify them with strong ladder correlations
dominant at short range. One such theory was applied by Green {\em et al}.
\cite{GNS1}, set up in part to understand angle-resolved inelastic
X-ray scattering off metallic films
\cite{GNS2}. Sec. V discusses two comprehensive theories of correlations:
the maximally coupled parquet model
\cite{HB, JLS, JLS1, JLS2, JLS3} and a simplification of it,
the induced-interaction approximation of Babu and Brown
\cite{BB,AB} suited to such systems as nuclear matter, liquid helium,
and extended later to low-density electron fluids
\cite{AGP}.
The summary is in Section VI.

Two Appendices follow the main body of the paper.
Appendix A reviews the third frequency-moment
sum rule in the electron gas as an example of how sum-rule validity,
albeit generic to stochastic Hamiltonian models, also requires discretion
when extracting the physical content of a reduced
correlation theory. Appendix B looks briefly towards possible
implications of the stochastic-Hamiltonian approach for nonperturbative
analyses of correlated systems. As recalled above, these
(in particular density-functional theory
\cite{dft1,dft2} and coupled-cluster analysis
\cite{cc1,cc2})
coexist with perturbative methods and likewise
derive from an exact originating Hamiltonian.
The potential for establishing explicit model structures for these
alternative and powerful formulations of the quantum many-body problem
could repay a thorough study.
 
\section{Stochastic Hamiltonian Models}

\subsection{Basic Formulation}

We begin with the standard Hamiltonian for a fermion system,
comprising a single-particle part and an interaction part
interacting via a pairwise potential:
\begin{eqnarray}
H
=&&
\sum_k \epsilon_k a^*_k a_k + H_i,
\cr
\cr
H_i
=&&
{1\over 2}
{\sum_{k_1 k_2 k_3 k_4}}\!\!\!\!\!\!'
{\langle k_1 k_2 | V | k_3 k_4 \rangle}
a^*_{k_1} a^*_{k_2} a_{k_3} a_{k_4};
~~~
\cr
\cr
&&
{\langle k_1 k_2 | V | k_3 k_4 \rangle}
\equiv
\delta_{s_1 s_4} \delta_{s_2 s_3} 
V({\bf k}_1 - {\bf k}_4).
\label{k01}
\end{eqnarray}
Notation is as follows. Index $k$ denotes state wavevector ${\bf k}$
and spin $s$ so $a^*_k$ is the creation operator in state $k$
and $a_k$ is the annihilation operator; both satisfy the usual
anticommutation relations. Here, the potential is spin-independent.
The summation ${\sum}'_{k_1 k_2 k_3 k_4}$ comes with the momentum
conservation restriction $k_1+ k_2 = k_3 + k_4$. In a uniform Coulomb
system with neutralizing background, the terms in $V({\bf 0})$ are
excluded.

The matrix element of the potential satisfies hermiticity and pairwise
exchange symmetry:
\begin{eqnarray}
{\langle k_4 k_3 | V | k_2 k_1 \rangle}
&=&
{\langle k_1 k_2 | V | k_3 k_4 \rangle}^*;
~~~
\cr
\cr
{\langle k_2 k_1 | V | k_4 k_3 \rangle}
&=&
{\langle k_1 k_2 | V | k_3 k_4 \rangle}.
\label{k00}
\end{eqnarray}

The Hamiltonian presented is assumed exact for the system of interest.
The first step in the Kraichnan construction is to posit a large number
$N$ of Hamiltonians identical to that of Eq. (\ref{k01}) but
whose fermion states are distinguishable. In other words, an additional
$N$-fold spin-like label is assigned to each system. We form the total
Hamiltonian for the assembly:
\begin{eqnarray}
{\cal H}_N
=&&
\sum^N_{n=1} \sum_k \epsilon_k a^{*(n)}_k a^{(n)}_k
+ {1\over 2} \sum^N_{n=1}
\cr
\cr
&& \times \!\!\!
{\sum_{k_1 k_2 k_3 k_4}}\!\!\!\!\!' ~
{\langle k_1 k_2 | V | k_3 k_4 \rangle}
~a^{*(n)}_{k_1} a^{*(n)}_{k_2} a^{(n)}_{k_3} a^{(n)}_{k_4};
\label{k03}
\end{eqnarray}
the additional superscript $n$ distinguishes the populations.

Next we map the assembly in Eq. (\ref{k03})
to a ``collective'' description. This is done by canonically
transforming the operators into a complementary set over the large,
but still finite, space $N$. For integer $\nu \leq N$ introduce
\begin{eqnarray}
a^{*[\nu]}_k
\equiv&&
N^{-1/2} \sum^N_{n=1} e^{2\pi i \nu n/N} a^{*(n)}_k
~~ {\rm and} ~~
\cr
\cr
a^{[\nu]}_k
\equiv&&
N^{-1/2} \sum^N_{n=1} e^{-2\pi i \nu n/N} a^{(n)}_k
\label{k04}
\end{eqnarray}
with the usual Fourier-series convention that sums of collective indices
$\nu$ are defined modulo $N$. The collective
operators of Eq. (\ref{k04}) satisfy the same anticommutation
relations as the original operators. The total Hamiltonian becomes
\begin{eqnarray}
{\cal H}_N
=&&
\sum^N_{\nu=1} \sum_k \epsilon_k a^{*[\nu]}_k a^{[\nu]}_k
+
{1\over 2N} {\sum_{k_1 k_2 k_3 k_4}}\!\!\!\!\!' ~~
\sum^N_{\nu_1 \nu_2 \nu_3 \nu_4}
\delta_{\nu_1+\nu_2, \nu_3+\nu_4}
\cr
\cr
&& {~~~ ~~~ ~~~ }
\times
~{\langle k_1 k_2 | V | k_3 k_4 \rangle}~
a^{*[\nu_1]}_{k_1} a^{*[\nu_2]}_{k_2} a^{[\nu_3]}_{k_3} a^{[\nu_4]}_{k_4}.
\label{k05}
\end{eqnarray}
 
\subsection{Stochastic Ansatz}

The ground is ready for Kraichnan's procedure. The object
described by Eq. (\ref{k05}) remains in every respect the
exact Hamiltonian, merely replicated $N$ times in distinguishable
but otherwise identical Hilbert spaces. Within its new collective
representation, however, it is possible to modify the interaction
by introducing couplings specifically tailored to enhance certain
classes of correlated expectation values, suppressing the remainder.
In the process the modified collective Hamiltonian retains
its functional properties.

All of the Hilbert-space machinery and the consequences
from the fundamental equation of motion
continues to apply to the collective Hamiltonian.
After ensemble averaging, those identities particularly
determined by analyticity of the expectation values will
survive averaging, since their causal structure is preserved.
More care is needed with any identities that
depend explicitly on completeness in Hilbert space, which
may not survive averaging. This is discussed in Appendix A.

Following Kraichnan we define restriction variables
$\varphi_{\nu_1 \nu_2|  \nu_3 \nu_4}$
to adjoin to the interaction potential. The ensemble interaction
Hamiltonian becomes
%
\begin{eqnarray}
{\cal H}_{i;N}
=&&
{1\over 2N} {\sum_{k_1 k_2 k_3 k_4}}\!\!\!\!\!' ~~ 
\sum^N_{\nu_1 \nu_2 \nu_3 \nu_4}
\delta_{\nu_1+\nu_2, \nu_3+\nu_4}
\cr
\cr
&& {~~~ ~~~ }
\times
\varphi_{\nu_1 \nu_2| \nu_3 \nu_4} ~
{\langle k_1 k_2 | V | k_3 k_4 \rangle}
a^{*[\nu_1]}_{k_1} a^{*[\nu_2]}_{k_2} a^{[\nu_3]}_{k_3} a^{[\nu_4]}_{k_4}.
\label{k06}
\end{eqnarray}
To maintain the hermiticity and label symmetry of $V$ itself, the
parameter $\varphi_{\nu_1 \nu_2| \nu_3 \nu_4}$,
must satisfy its corresponding form of Eq. (\ref{k00}),
noting that the collective creation and annihilation operators
in Eq. (\ref{k06}) bind together the collective labels
$\nu_j$ and system basis state labels $k_j$ for $j = 1, 2, 3, 4$.

The properties of $V$ shared by $\varphi$ are crucial to the entire exposition.
They establish the microscopic equivalence of the Kraichnan procedure
to the Baym-Kadanoff rules \cite{KB,GB} for constructing
conserving, or ``$\Phi$-derivable'', models of the interacting
free-energy functional.
Given these constraints,
the introduced variable will couple the formerly independent
Hamiltonian components in any way one wishes
without upsetting the analytic
structure of the $N$-fold system. In particular, they can be assigned
randomly determined values.

When the choice of $\varphi$ is not random, the consequences
are immediately reflected in the $N$-fold Hamiltonian.
When the choice is random the Hamiltonian, altered in this way,
is to be embedded within a still larger ensemble.
Each member of this super-collection has an identical form
in terms of the restriction parameters but each is
characterized by its own specific set of stochastic values.
Depending on the restrictions' internal structure,
certain products of them will cancel within the
diagrammatic expansion of the ground-state energy.
These terms are the subset of correlations designed to survive
the final ensemble averaging over the assigned value of $\varphi$.
All other terms will tend to interfere
destructively, to be quenched in the ensemble average.

Thus Kraichnan's calculational philosophy is exactly
that of Bohm and Pines' RPA, albeit far more flexible.
In principle, such a construct is able to generate models
with a vast range of selected perturbation terms -- to all orders when
required -- naturally dictated by the physical context
to be captured. The selection is expressed through the particular
restrictions imposed via $\varphi$.

Every such implementation is a truncation to the complete
many-body problem, although the truncation can be very sophisticated.
Throughout the reduction, each model still possesses a well-defined
Hamiltonian ensemble respecting -- in its own reduced fashion -- all
of the relevant analytic identities, and their inter-relationships,
inherent in the exact description. Quantitatively, of course, the
changes might be drastic while, qualitatively, the generic behavior
and development of the system under its Hamiltonian will always
apply. The power of the approach consists precisely in this.

Before reviewing classic examples of reduced Hamiltonians (including
from the original Kraichnan study) and going on to different and more
comprehensive correlation models, we recapitulate the procedural logic.

\begin{itemize}
\item Conceptualize a sufficiently large number $N$ of dynamically identical,
but distinguishable, copies of an exact Hamiltonian. Each retains the same
interaction potential but all copies are mutually uncoupled. This enlarged
Hamiltonian exhibits physics completely identical to any one of the
embedded copies of the exact system.

\item Fourier transform the state operators of each copy to a new set of
operators for a coherent pseudo-collective superposition of the $N$ systems.
The transformation generates a new set of indices, formally analogous
to the wavevector states in reciprocal space.

\item For the pseudo-collective description, introduce a set of
restriction factors labeled by the new collective indices. Adjoin
each factor to the basic (unaffected) interaction potential.

\item The functional form of the factor must satisfy the same label
symmetries as does the potential with respect to its state labels.
This is equivalent to the Baym-Kadanoff rules
\cite{KB,GB} and preserves the modified Hamiltonian as a hermitian operator,
with the all the canonical identities also preserved.

\item The $N$ copies are now interlinked via the introduced factors,
and these couplings can be assigned values within any desired protocol;
in particular, they may be defined stochastically.

\item
Finally, ensemble-average the system over the distribution governing the
coupling factors. Since each parametrized $N$-fold Hamiltonian retains
its fundamental properties, analytic relations
among expectation values will be preserved
functionally -- but not usually numerically -- after averaging. 
\end{itemize}

The central element is to have made sure that all such model
Hamiltonians retain the analytic characteristics of the original
physical Hamiltonian. To establish quantitative
results from the model, one simply follows the same theoretical
steps applicable to the exact system.
This is of enormous help in confirming microscopic conservation
for any model, notably through the dynamical sum rules
that condition its response and fluctuation structure
\cite{DP2}.

Whether or not the resulting numbers are adequate to the physical situation
one wants to analyze, depends strictly on how the ensemble parameters
$\varphi$ are chosen, just as in the more directly intuitive and synthetic
$\Phi$-derivable approach
\cite{KB,GB}. Nevertheless, certain
basic inner relationships such as sum rules among derived quantities
remain valid throughout.

\section{Instances of Model Hamiltonians}

We preface the later extension of Kraichnan's construction to other
many-particle model systems by reviewing the classic examples.
We first introduce the RPA model before revisiting the classic
formulations first analyzed by Kraichnan.

\subsection{Random-Phase Approximation}

The Bohm-Pines RPA can be obtained by defining its variable as
\begin{equation}
  \varphi^{\rm RPA}_{\nu_1\nu_2| \nu_3 \nu_4}
= \delta_{\nu_1 \nu_4} \delta_{\nu_2 \nu_3}
\label{k07.0}
\end{equation}
This non-random assignment fulfils the symmetries of Eq. (\ref{k00}).
Its effect is illustrated in Fig. 1 via its contributions to the
ground-state correlation-energy functional.
In a many-body system, only ``linked'' diagrams, namely those consisting
of a single diagrammatic unit, represent valid contributions to the
correlation energy
\cite{GB}. The object that results from the prescription
in Eq. (\ref{k07.0}) is just the direct Hartree
(or mean-field) correlation energy,
Fig. 1(b). All other would-be contributions to
higher order in $V\varphi$ that do not vanish identically
turn out to be unlinked in the summation over
indices, and therefore do not enter into the
canonical correlation-energy functional.

%
\centerline{
\includegraphics[height=4truecm]{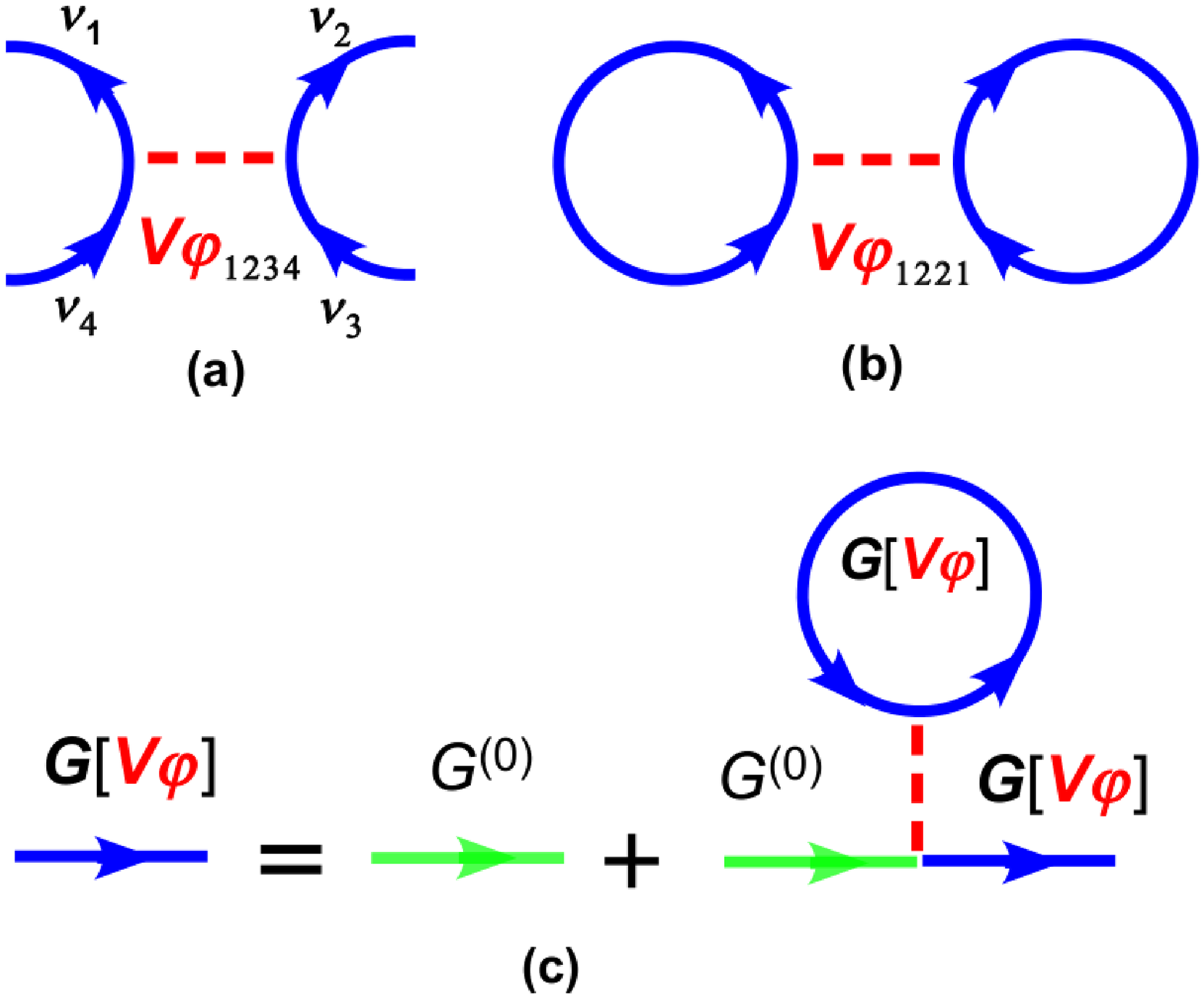}
}
{{\bf FIG. 1.} {\small Structure of the random-phase approximation,
or RPA. Continuous lines: incoming/outgoing particle. Broken lines:
interaction. The restriction parameter $\varphi^{\rm RPA}$ of equation
(\ref{k07.0}) generates the coupling shown in (a). Successive
interaction terms in the Hamiltonian cannot interlink through
this parameter. Only the single linked diagram shown in (b)
survives to define the ground-state correlation energy. Given the
Hamiltonian, the Dyson equation, symbolized in (c), can be
set up directly for the RPA single-particle Green function
$G[V\varphi]$ starting from the noninteracting Green function $G^{(0)}$. 
Although the RPA correlation energy has the simplest possible
structure of any model, the high level of self-consistency is
evident through the structure of the Dyson equation (c).}}
\vskip 0.15cm

When the Hamiltonian is augmented with an external perturbation,
the associated Heisenberg equation of motion
\cite{mahan} leads systematically
to both one-body and two-body dynamical Green functions, or propagators.
These contain the necessary information for computing those response
functions that can be compared with experimental measurements.
Figure 1(c) shows the prototypical Dyson integral equation
\cite{mahan} for the one-body propagator within the RPA entering
into the energy functional
of Fig. 1(b). Notwithstanding the structural simplicity of the
random-phase approximation, this reveals the high degree of internal
nesting that lies implicitly concealed within it, as with any nontrivial
theory of many-body correlations.

\subsection{Hartree-Fock}

The next simplest model is Hartree-Fock (HF), which introduces
the primary exchange corrections to the RPA. In place of 
Kraichnan's own choice for selecting the Hartree-Fock Hamiltonian,
we adapt the same Ansatz as for RPA after antisymmetrizing
the original pair interaction following Nozi\`eres
\cite{nozi}. This is done by exchanging one
pair of incoming or outgoing indices, say $3 \leftrightarrow 4$
for definiteness, and using anticommutation to replace
${\langle k_1 k_2 | V | k_3 k_4 \rangle}$ with
\begin{equation}
{\langle k_1 k_2 | {\overline V} | k_3 k_4 \rangle}
\equiv {1\over 2}
{\Bigl( {\langle k_1 k_2 | V | k_3 k_4 \rangle}
       -{\langle k_1 k_2 | V | k_4 k_3 \rangle} \Bigr)}
~~~~
\label{k07.1}
\end{equation}
in the full Hamiltonian. It makes no change to the physics,
but means that the RPA Ansatz Eq. (\ref{k07.0})
also covers the exchange vertex as in Fig. 2 (a).
The outcome is the Hartree correlation energy of Fig. 1(b) once again,
now accompanied by its Fock exchange counterpart. In a Coulomb system,
the long-ranged effects remain subsumed under the Hartree structure.
At shorter range, comparable to the system's Fermi wavelength, the Fock
term corrects for Pauli repulsion, which is absent from RPA
causing it to overestimate the Coulomb energy.
\centerline{
\includegraphics[height=4truecm]{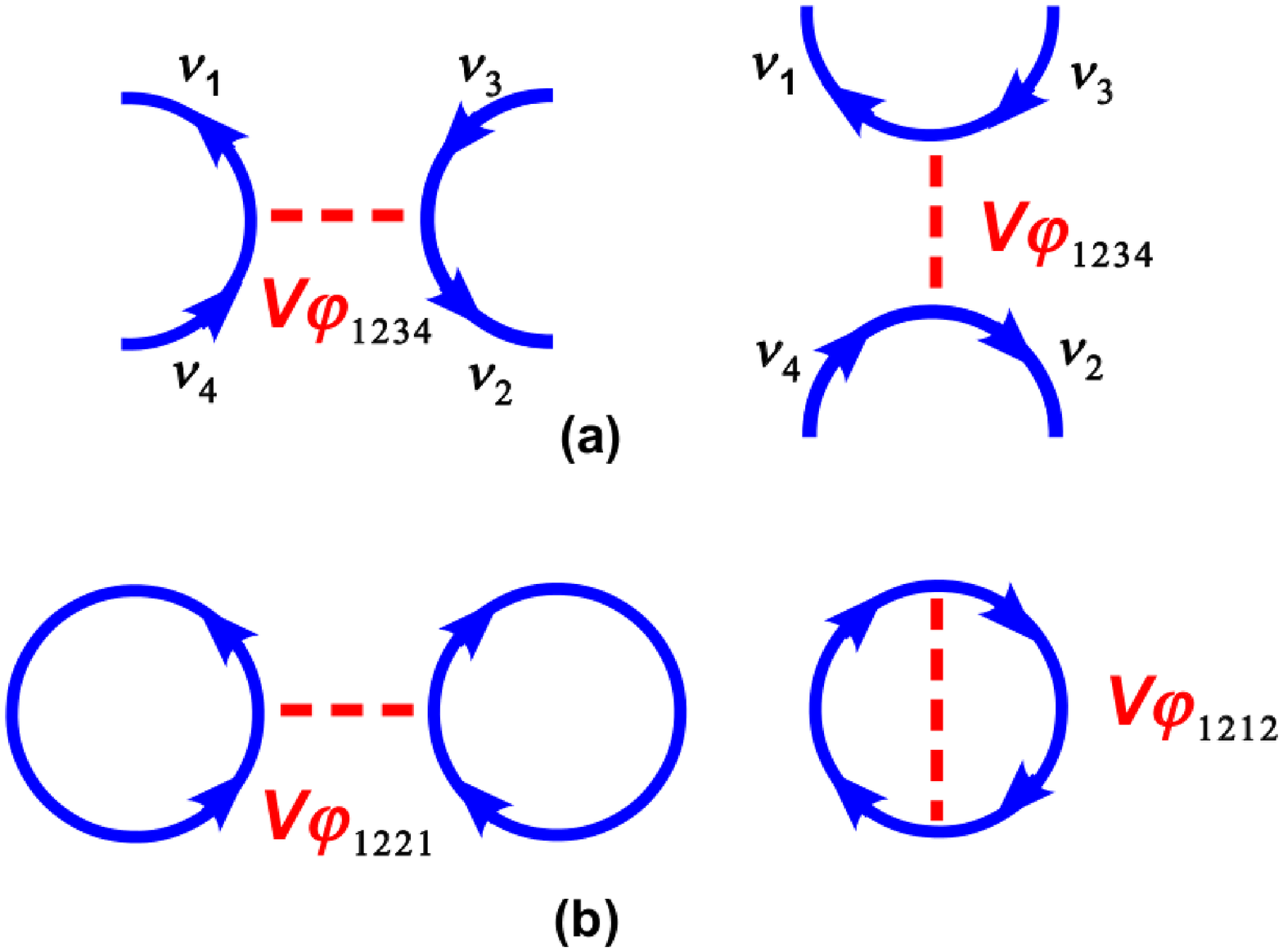}
}
{{\bf FIG. 2.} {\small Correlation diagrams
associated with the model Hamiltonian carrying the
Kraichnan RPA factor $\varphi^{\rm RPA}_{\nu_1\nu_2| \nu_1 \nu_2}$ -- see
Eq. (\ref{k07.0}) in text -- now with exchange
explicitly incorporated in the interaction potential.
The allowed topological possibilities for the two-body vertex
are shown in (a).
The only linked diagrams to survive the trace over $\varphi^{\rm RPA}$
are those of (b), exhibiting the standard Hartree
and Fock-exchange correlation energy terms.
With the potential antisymmetrized, each of the vertices of (a) will
contribute half of the total direct and exchange terms of (b).
(Combinatorial weightings for the
correlation diagrams will not be shown; they are identical
to the standard derivation of the ground-state functional
\cite{KB}.)
}}
\subsection{Shielded Interaction}

The first truly stochastic Ansatz introduced by Kraichnan regenerates
the shielded-interaction, or ``ring'', approximation
\cite{KB} closely related to but richer than the pure
RPA and HF. In a system with long-ranged potential, it
provides the leading short-range correlation corrections to the
screening properties of the system. For the ring model,
the restricting factors are defined in terms of a uniform
random distribution of phase angles so that
\begin{eqnarray}
&&
\varphi^{(r)}_{\nu_1 \nu_2 \nu_3 \nu_4}
\equiv
\exp[\pi i(\zeta_{\nu_1 \nu_4} + \zeta_{\nu_2 \nu_3})];
~\zeta_{\nu\nu'} \in [-1,1];
\cr
\cr
&&
\zeta_{\nu'\nu}
=
-\zeta_{\nu\nu'}.   
\label{k10}
\end{eqnarray}
\centerline{
\includegraphics[height=4truecm]{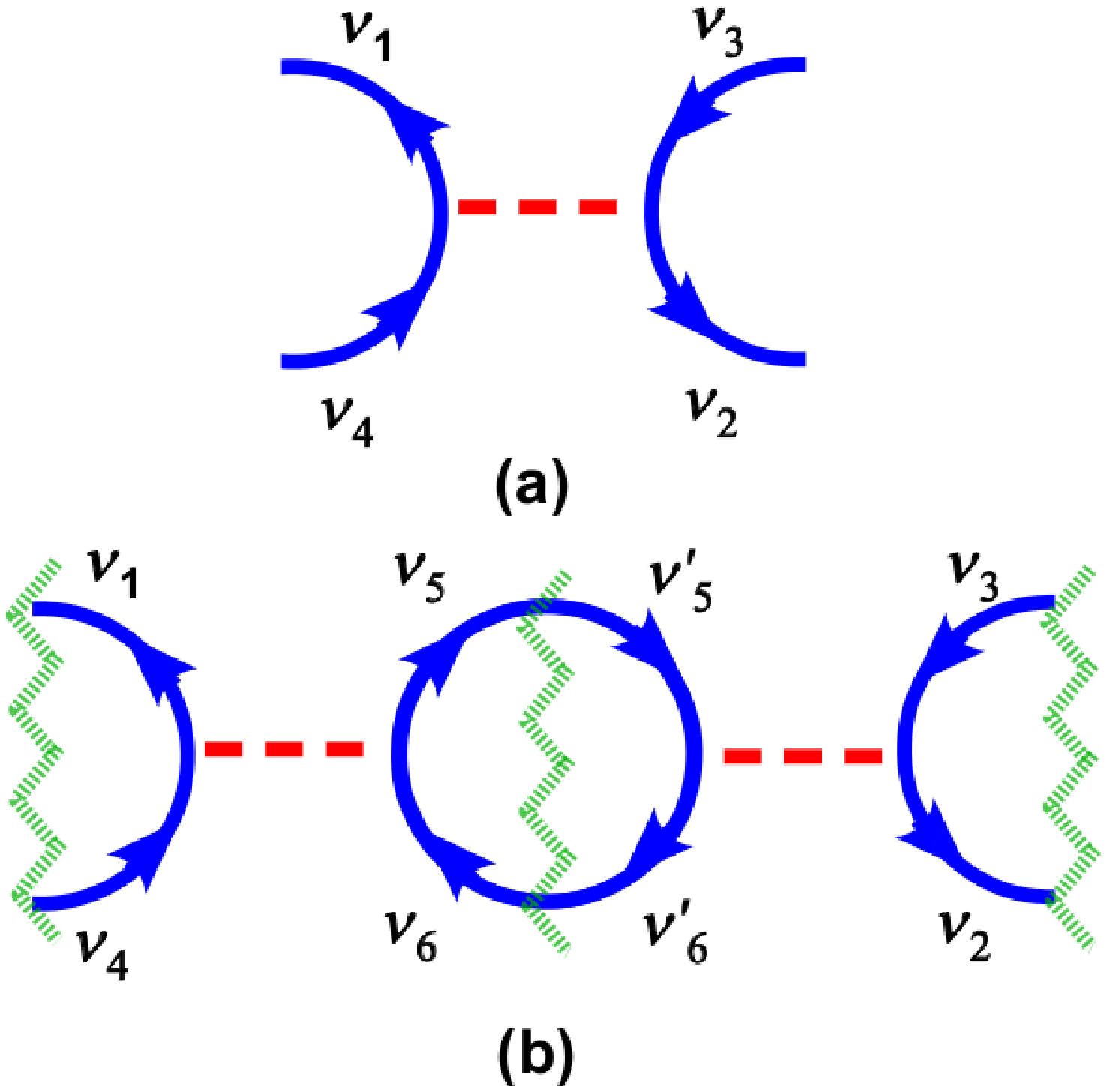}
}
{{\bf FIG. 3.} {\small Allowed vertex contributions
for the RPA stochastic Hamiltonian model; see Eq. (\ref{k10})
in the text.
(a). Lowest-order (Hartree) term.
(b). All other surviving terms can only adopt a repeated chain-link
topology. The outer open lines of all terms may link in two different
ways, leading either to open chains or simple rings.
}}
\vskip 0.15cm
The phase reverses when the roles of an outgoing and
incoming pair of lines reverse (particle $\leftrightarrow$ hole).
Kraichnan's choice of a phase Ansatz always follows a possible
action of the local particle operators (creation and annihilation) inside
the diagrammatic structures one wants to highlight.
Here, $\zeta_{\nu\nu} \equiv 0$ for a self-closing line.

Consider Fig. 3(b) to second order in the interaction.
Closing the intermediate lines enforces equality of
the intermediate pairs $\nu_5, \nu'_5$ and $\nu'_6, \nu_6$
to form an elementary particle-hole polarization
``bubble''. The concatenation
$\varphi_{\nu_1 \nu_5| \nu_6\nu_4} \varphi_{\nu_6 \nu_2| \nu_3 \nu_5}$
then leads to the net phase
\begin{equation}
(\zeta_{14} + \zeta_{56}) + (\zeta_{65} + \zeta_{23})
= \zeta_{14} + \zeta_{23}.
\label{k11}
\end{equation}
It is clear that the cancellation observed in Eq. (\ref{k11})
persists to all orders in the ground-state diagram expansion.
This secures the survival of the chain-like terms.
For any other topology the phases will not cancel and will
thus be suppressed in the trace over the stochastic ensemble.

An illustrative example of correlated terms surviving the
ensemble average for the ring model appears in Fig. 4 where
we display its associated density-density response,
or polarization, function
\cite{DP2}.
In the momentum-frequency domain the characteristically screened
interaction, as defined in Fig. 4(c), pairs the lowest-order
polarization $\chi(q,\omega)$ with
the bare potential $V(q)$. The summation runs to all orders,
but {\em only} for those components allowed by the parameter
Eqs. (\ref{k10}) and (\ref{k11}).
%
\centerline{
\includegraphics[height=5truecm]{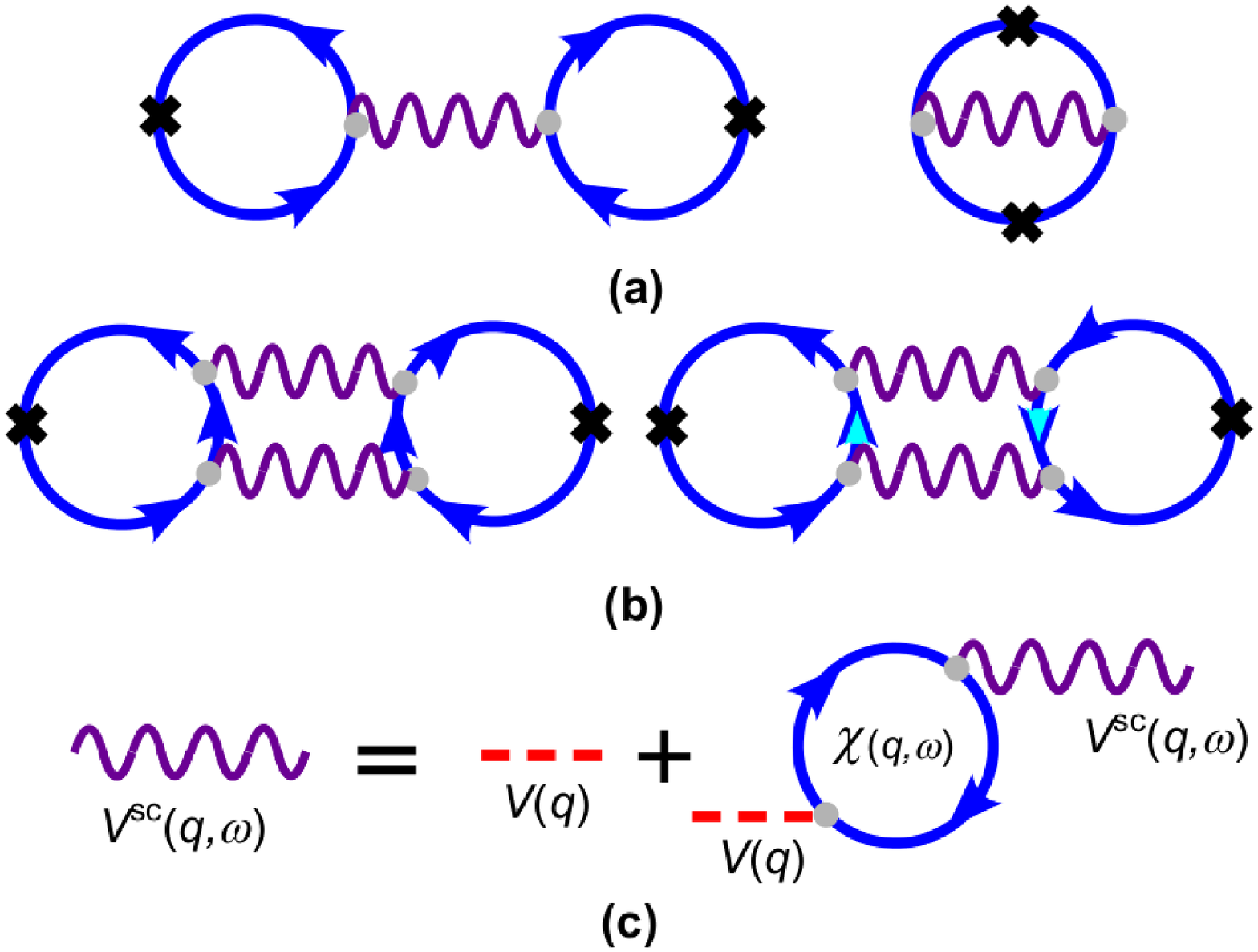}
}
{{\bf FIG. 4.} {\small Topology of the 
correlation corrections to the density-density response function
in the stochastic ring model. Crosses denote coupling to an
external perturbation of the density.
Screened Hartree-Fock contributions appear
in (a) while (b) shows a new pair of correlated contributions
necessarily appearing in the response to satisfy
microscopic conservation at the two-body level
\cite{KB,dubois}; note that they are
functionally distinguished by the mutual orientation of their
attached loops: in the left-hand term the screened potentials are
connected by two particle lines while in the right-hand term they
are connected by a particle and a hole. The intermediate propagators
for these processes act differently. In (c) the equation for
the self-consistent screened interaction of (a) and (b) is defined.
The object $\chi(q,\omega)$ is the leading term in the total
polarization response.
}}

\subsection{Particle-Particle Ladder}

The ring model builds upon the RPA/HF by incorporating
the next level of screening corrections at finite range,
but does not do well for shorter-ranged interactions with a
hard core, such as nucleons or neutral atomic fluids
where the extreme degree of local repulsion between particles
invalidates the finite-order Born approximation
\cite{BB}.
Accounting for hard-core effects requires the
ladder approximation of Brueckner
\cite{bru}, designed to accommodate the extreme distortion
in the pair correlation function from the interaction at close range.
The aim is to incorporate the full Born series for
two-particle scattering in the interacting medium,
using the Bethe-Salpeter equation
\cite{nozi}.
%
\centerline{
\includegraphics[height=3truecm]{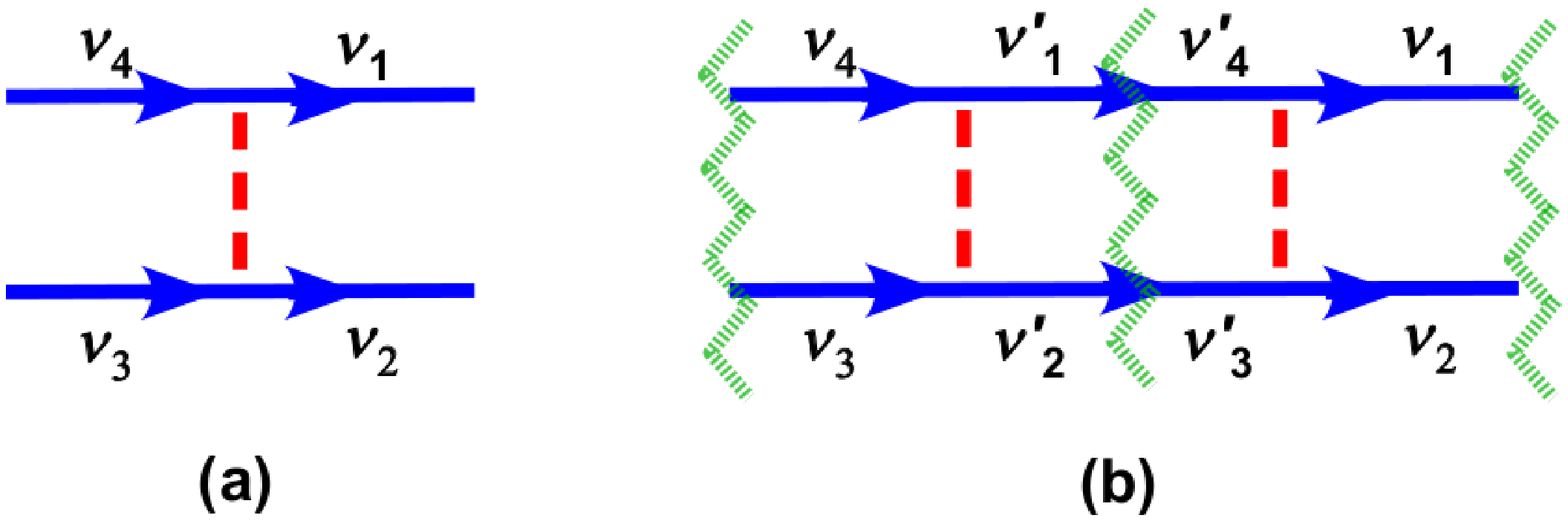}
}
{{\bf FIG. 5.} {\small Repeated sequences, or ladders,
of particle-particle dynamical correlations with the stochastic
restriction parameter of Eq. (\ref{k12}). The elementary
scattering term is in (a) while in (b) all stochastic phases of inner
propagator pairs cancel in all higher orders.
}}
\vskip 0.15cm

Following Kraichnan, the ladder-model Hamiltonian
is defined by the restriction parameter
\begin{eqnarray}
&&\varphi^{(pp)}_{\nu_1 \nu_2 | \nu_3 \nu_4}
\equiv
\exp[\pi i(\xi_{\nu_1 \nu_2} - \xi_{\nu_3 \nu_4})];~
\xi_{\nu \nu'} \in [-1,1];~
\cr
\cr
&&
\xi_{\nu' \nu}
=
\xi_{\nu \nu'}.   
\label{k12}
\end{eqnarray}

\noindent
While $\varphi^{(r)}$ for the ring model favors particle-hole pair
propagation via polarization bubbles, $\varphi^{(pp)}$ for the ladder
approximation privileges two-particle propagation mediated not
by the bare interaction but by its complete pairwise scattering matrix.
This is shown in Fig. 5. When the restriction factors are concatenated
as with Eq. (\ref{k11}) of the ring model, this time the pattern
for the sum of phases is
\begin{equation}
  (\xi_{\nu_1\nu_2} - \xi_{\nu'_3 \nu'_4})
+ (\xi_{\nu'_1\nu'_2} - \xi_{\nu_3 \nu_4})
= \xi_{\nu_1\nu_2} - \xi_{\nu_3 \nu_4}
\label{k13}
\end{equation}
since the algebra of creation-annihilation pairing now forces
$\nu'_1 = \nu'_4$ and $\nu'_2 = \nu'_3$. From Eq. (\ref{k13})
the same cancellation obtains
under exchange $\nu'_1 = \nu'_3$ and $\nu'_2 = \nu'_4$, leading
to a contribution analogous to the Fock term in the
ground-state correlation energy. Figure 6 illustrates
the polarization corrections expected within the ladder approximation.
\centerline{
\includegraphics[height=6truecm]{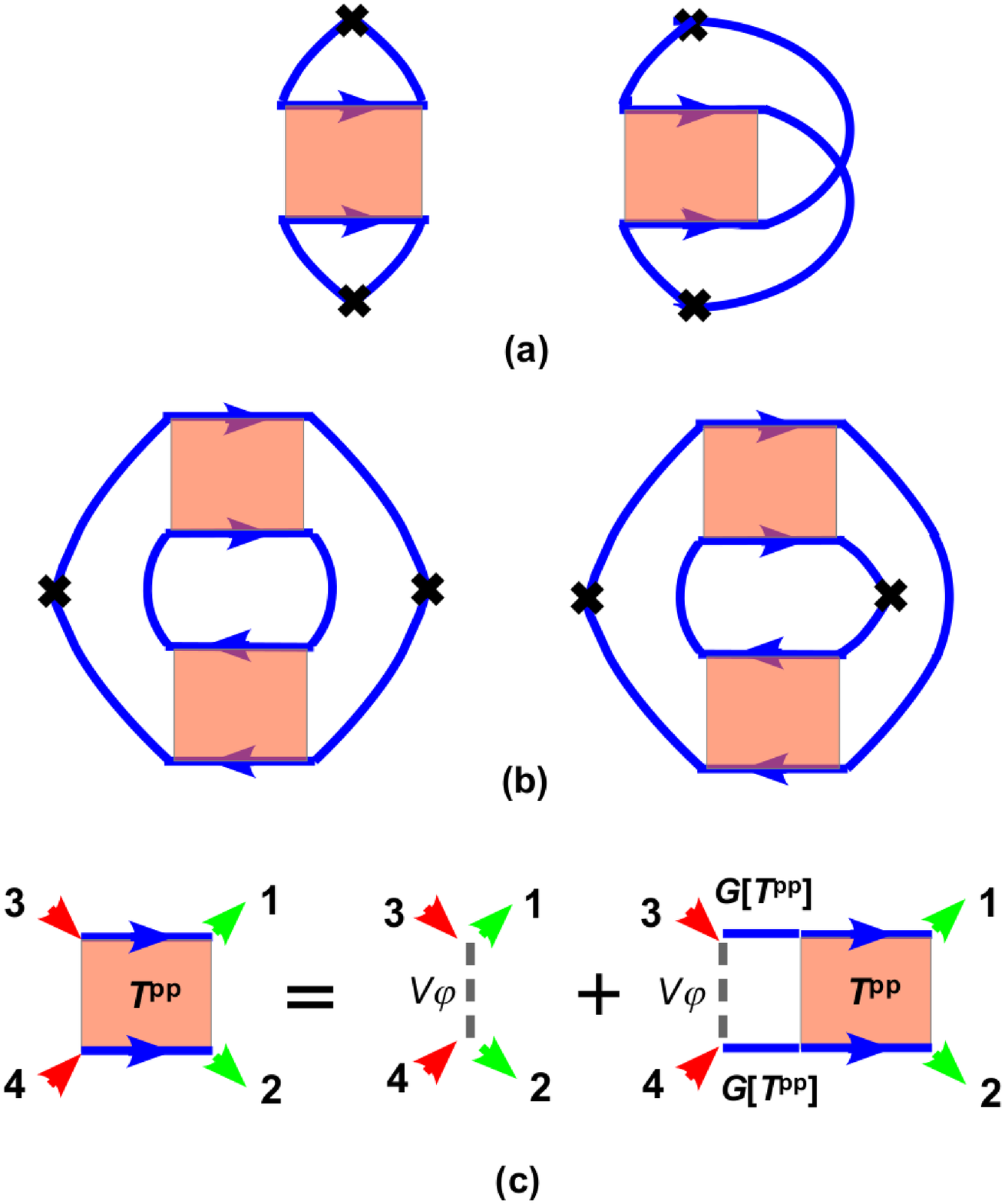}
}
\vskip 0.25cm
{{\bf FIG. 6.} {\small Irreducible corrections to the polarization
response in the
particle-particle ladder approximation. These are obtained,
as for the ring model, by a standard variational procedure starting
from many-body ground-state functional.
Diagrams in (a) carry a single particle-particle ladder-scattering
vertex. Note that the exchange term on the right of (a) is already
accounted for within the term on the left and is presented merely
to bring out the internal topology of this contribution. Diagrams
in (b) display the two possibilities (a consequence of conservation
\cite{KB}) by which the particle-particle amplitude
also mediates intermediate particle-hole processes.
The Bethe-Salpeter equation for the scattering vertex,
or particle-particle $T$-matrix $T^{pp}$, is schematized in (c).
The single-particle propagator is similarly self-consistently
defined by $T^{pp}$ through the Dyson equation for the model.
}}
\vskip 0.15cm

\subsection{Particle-Hole Ladder}

The particle-hole ladder model extends the exchange
structure of Hartree-Fock in the way that the ring model does for
the direct random-phase approximation. This particular scattering
channel will be needed in Section V.
With a slight change to the particle-particle
mechanism, we generate its particle-hole analog. Consider
\begin{eqnarray}
&&\varphi^{(ph)}_{\nu_1 \nu_2 | \nu_3 \nu_4}
\equiv
\exp[\pi i(\vartheta_{\nu_1 \nu_3} + \vartheta_{\nu_2 \nu_4})];~
\vartheta_{\nu \nu'} \in [-1,1];
\cr
\cr
&&\vartheta_{\nu' \nu}
=
-\vartheta_{\nu \nu'}.   
\label{k20}
\end{eqnarray}
\vskip 0.25cm
\centerline{
\includegraphics[height=4.5truecm]{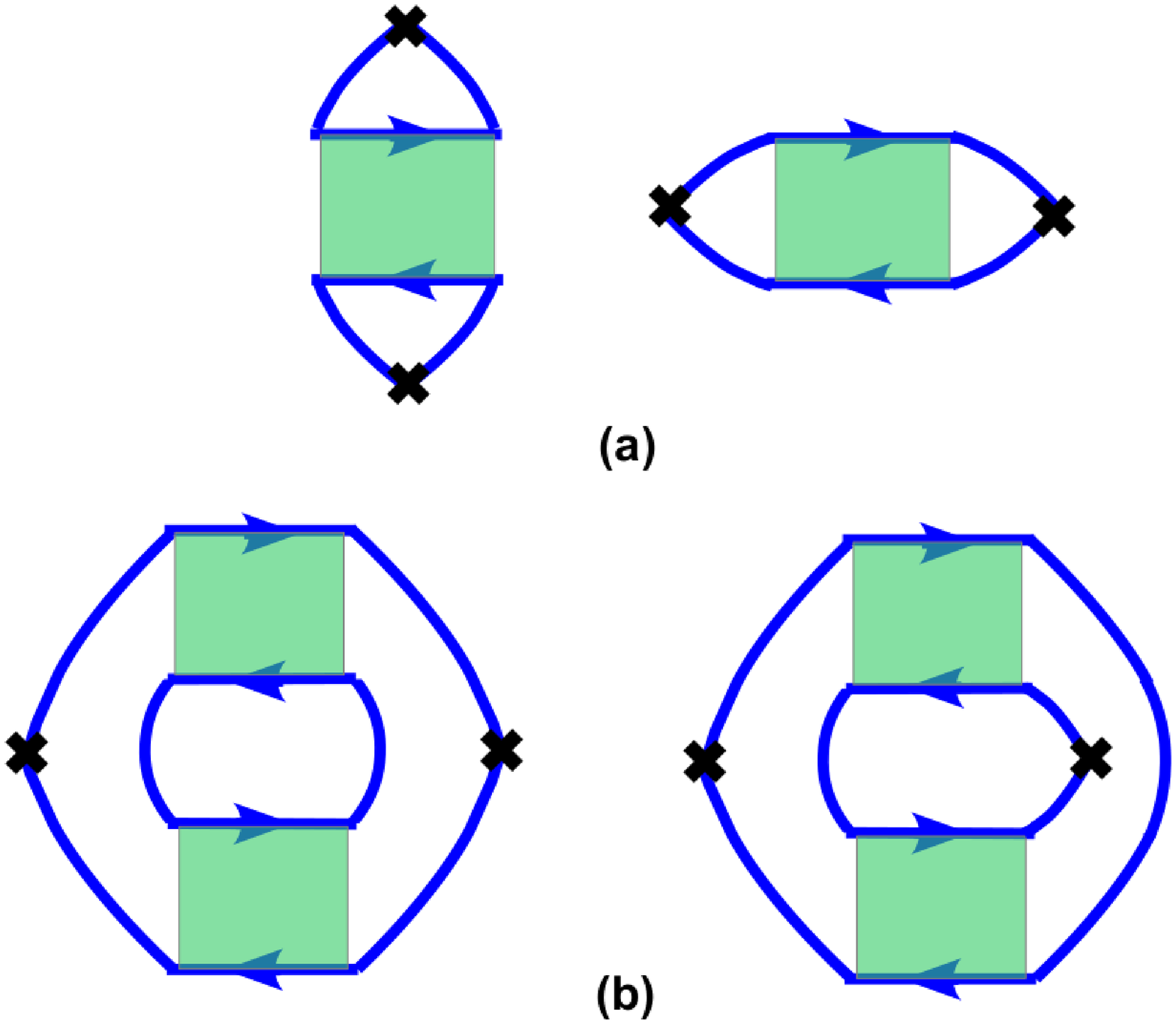}
}
{{\bf FIG. 7.} {\small Irreducible polarization corrections for the
particle-hole ladder interaction associated with Eq. (\ref{k20}).
Diagrams in (a) carry a single
ladder-scattering vertex. Diagrams in (b) display the two additional
topological possibilities in which the vertex also mediates intermediate
particle-hole propagation within the polarization function.
The Bethe-Salpeter equation for the particle-hole ladder is essentially
that for particle-particle scattering except that one set of
particle lines is reversed and exchange is excluded.
}}
\vskip 0.25cm

Particle-hole pairings in the elementary interaction vertex
are coupled as if the hole were a particle; reversing the roles
in the pair reverses their phase, as one would expect.
The leading corrections to the polarization for
the particle-hole ladder model,
analogous to those of Fig. 6 for the particle-particle ladder,
are in Fig. 7 including the terms of 7(b) of second order in the total
scattering amplitude and required by conservation.

This ends the review of the Hamiltonian formulations first presented
by Kraichnan for standard correlation models. We have added
the random-phase approximation in its own right (otherwise subsumed
by Kraichnan under his Hartree-Fock prescription) as well as
the particle-hole ladder. In the following section we explore more
comprehensive Hamiltonian models.

\section{Extensions of Kraichnan's Method}

\subsection{Systematic Removal of Correlations: Screened Hamiltonian}

We start by posing the problem of how an interacting Hamiltonian
may change when certain components are removed selectively, as one
would do for closer analysis of the remnant correlations.
As an example we isolate the RPA Hamiltonian from the exact
description. There is no loss of physical content in
recasting Eq. (\ref{k05}) as
\begin{eqnarray}
&&{\cal H}
=
{\cal H}^{\rm RPA} +
{1\over 2N}  \sum_{\ell_1 \ell_2 \ell_3 \ell_4}
\delta_{\ell_1+\ell_2, \ell_3+\ell_4}
\cr
&& %
{~~~ ~~~ } 
\times
(1 - \varphi^{\rm RPA}_{\nu_1 \nu_2 | \nu_3 \nu_4})
~ {\langle k_1 k_2 | V | k_3 k_4 \rangle}
a^*_{\ell_1} a^*_{\ell_2} a_{\ell_3} a_{\ell_4};
\cr
\cr
&&{\cal H}^{\rm RPA}
=
\sum_{\ell} \epsilon_{k} a^*_{\ell} a_{\ell}
\cr
&&
{~~~ ~~~ }
+ {1\over 2N} {\sum_{\ell_1 \ell_2 \ell_3 \ell_4}}\!\!\!\!' ~~
\varphi^{\rm RPA}_{\nu_1 \nu_2| \nu_3 \nu_4} ~
{\langle k_1 k_2 | V | k_3 k_4 \rangle}
a^*_{\ell_1} a^*_{\ell_2} a_{\ell_3} a_{\ell_4}.
\label{k15}
\end{eqnarray}
To streamline the notation from now on, in the second expression
of Eq. (\ref{k15})
we have grouped the joint variables $\{k, \nu\}$
in $a^{[\nu]}_k$ into one symbol $\ell$ so $a^{[\nu]}_k \equiv a_{\ell}$.
Summations over $\ell$ encompass summations over both $k$ and $\nu$;
Kronecker deltas are now products of those in $k$s and$\nu$s and,
again, ${\sum}'_{\ell_1 \ell_2 \ell_3 \ell_4}$ is under the
constraint $\ell_1+\ell_2=\ell_3+\ell_4$ standing in for
$k_1+k_2=k_3+k_4$ and $\nu_1+\nu_2=\nu_3+\nu_4$ .

\noindent
So far nothing has changed; nor is there loss of any formal attribute
on introducing the reduced RPA-free version
\begin{eqnarray}
{\cal H}^{\rm sc}[\psi]
&\equiv&
\sum_{\ell} \epsilon_{k} a^*_{\ell} a_{\ell}
+ {1\over 2N} {\sum_{\ell_1 \ell_2 \ell_3 \ell_4}}\!\!\!\!\!' ~~ 
(1 - \varphi^{\rm RPA}_{\nu_1 \nu_2| \nu_3 \nu_4})
\cr
\cr
&&
\times
\psi_{\nu_1 \nu_2| \nu_3 \nu_4}~
{\langle k_1 k_2 | V | k_3 k_4 \rangle}
a^*_{\ell_1} a^*_{\ell_2} a_{\ell_3} a_{\ell_4}
\label{k16}
\end{eqnarray}
as long as the restriction parameter $\psi$ has the
symmetries required by Eq. (\ref{k00}).

The object in Eq. (\ref{k16}) represents all the
correlations of physical interest {\em except} for the
collective plasmon mode
\cite{DP2}.
In the language of Bohm and Pines this is the Hamiltonian for the
screened assembly: the part responsible for the near-field
dynamics experienced by a test particle immersed in the system.

First and foremost the screened Hamiltonian ${\cal H}^{\rm sc}$
has experimental relevance to metallic electron
systems in the normal state, since an external magnetic field will couple
to the spin density but not to the total charge density. In this
situation RPA screening does not contribute.

Besides this essential practical application, the theoretical
relevance of separating out the random-phase part is for the
sum rules. We illustrate the case of the $f$-sum rule, a familiar
identity expressing particle and energy conservation. Its proof
(see for example Nozi\`eres
\cite{nozi}) relies on the fact that the time-dependent operator
in the Heisenberg picture
\cite{mahan},
\[
\rho_{\ell \ell'}(t,t') \equiv a^*(t)_{\ell} a_{\ell'}(t'), 
\]
commutes with any pairwise interaction Hamiltonian -- exact or
reduced -- as long as the same label symmetries of the interaction are
satisfied both in physical and in Kraichnan's pseudo-collective spaces.

The $f$-sum rule connects the net energy absorbed from an external
perturbation to the energy distribution among the available
excitations of the system. In the classic case of the uniform
electron gas at zero temperature, it states (adopting units in which
$\hbar$ and the free electron mass are set to one)
\begin{eqnarray}
\int^{\infty}_{-\infty} {d\omega\over 2\pi}
\omega S(q, \omega)
= {q^2\over 2} n,
\label{k17}
\end{eqnarray}
in which $q$ is the momentum
transferred by the perturbation and $n$ is the electron density.
The dynamic structure factor
$S(q,\omega)$ is the density of states for all the system's
excited modes at momentum-energy $(q,\omega)$;
it is the negative imaginary part of the total dynamic polarization
$\chi(q,\omega)$, 
including the contribution from the collective plasmon mode.

Proof of the $f$-sum rule follows from the dynamical equations
for $\rho_{\ell \ell'}$ governed by the Hamiltonian. The rule
asserts that, no matter how the absorbed energy is redistributed
throughout the perturbed system (in more or less intricate ways),
in sum it is conserved and must account for the
energy gained per particle. The question is: does the electron
gas have an analogous rule when the dominant plasma mode is
``removed'' in a sense to be made precise?

The answer to the above is yes. This is almost obvious, since
the right-hand side of Eq. (\ref{k17}) has no dependence
on the interaction (and consequently is insensitive to all internal
correlations and all modifications to the potential that do
not alter its symmetries). In this form the rule is known commonly
as the conductivity sum rule.

One knows already that any canonical derivation
for the full Hamiltonian, for instance the $f$-sum rule, will be
valid for a reduced Hamiltonian. The logical form of such a proof,
once given for the exact case, does not care about the nature of any
appropriate reduction. Accordingly, let $S^{\rm sc}(q, \omega)$ be the
dynamic structure factor appropriate to ${\cal H}^{\rm sc}[\psi]$
of Eq. (\ref{k16}). If we now perturb this system, conservation
nevertheless applies and we obtain
\begin{eqnarray}
\int^{\infty}_{-\infty} {d\omega\over 2\pi}
\omega S^{\rm sc}(q, \omega)
= {q^2\over 2} n,
\label{k18}
\end{eqnarray}
or the conductivity sum rule, with essentially zero effort.

The Hamiltonian system ${\cal H}^{\rm sc}[\psi\!=\!1]$ preserves
all non-RPA contributions to the true ground state.
This is because its one-body propagators are unchanged by screening
as the Hartree mean-field term in the self-energy
\cite{mahan} is always canceled by local charge neutrality.
For non-uniform Coulomb systems this is not true in general,
but in the uniform situation the polarization $\chi^{\rm sc}(q,\omega)$,
whose imaginary part is $-S^{\rm sc}(q, \omega)$, contains only the
``proper'' correlations for the original system; that is, all
those that are not RPA. In that sense the
system becomes formally shielded from its long-range physics.
The consequent ability to validate sum-rule consistency
for any screened reduced model is of central importance;
while Eq. (\ref{k18}) then conveys no additional physical information,
it does provide an essential numerical test in implementing models
of the uniform electron fluid.

\subsection{Systematic Addition of Correlations: Ring-plus-Ladder Model}

In the previous Section we discussed two paradigms: the ring model,
which improves upon Hartree-Fock by including some shorter-ranged
correlations from the screened interaction (RPA, essentially),
and the particle-particle ladder model to treat strong short-ranged
effects beyond exchange. A combination of both was implemented by
Green, Neilson and Szyma\'nski
\cite{GNS1} for the electron gas to interpolate
between dominant long-range Coulomb screening and the short-range
Coulomb correlations expected to prevail at wavelengths
accessible in high-energy X-ray scattering
\cite{GNS2}.

The long-range-with-short-range interpolation was built bottom-up,
as it were, by isolating its physically dominant diagrams, the rings
and ladders of Figs. 4 and 6, out of the expansion of the
exact ground-state correlation energy.
These terms were duly symmetrized to make sure that they obeyed the
Baym-Kadanoff criteria for conserving, or $\Phi$-derivable, approximations
\cite{KB,GB}.

Typical of $\Phi$-derivable theories, the ring-plus-ladder model
was set up without a Hamiltonian,
rendering subsidiary derivations more burdensome than they might
have been. Here we present a stochastic Hamiltonian for
the Green {\em et al}. prescription:
\begin{eqnarray}
&&
{\!\!\!\!\!\! } {\!\!\!\!\!\! } {\!\!\!\!\!\! }
{\cal H}^{\rm GNS}
\equiv
\sum_{\ell} \epsilon_k a^*_{\ell} a_{\ell}
+ {1\over 2N} 
\!{\sum_{\ell_1 \ell_2 \ell_3 \ell_4}}\!\!\!\!' ~~
\!\!\!
\varphi^{\rm GNS}_{\nu_1 \nu_2| \nu_3 \nu_4}
{\langle k_1 k_2 | V | k_3 k_4 \rangle}
a^*_{\ell_1} a^*_{\ell_2} a_{\ell_3} a_{\ell_4};
\cr
\cr
&&
{\!\!\!\!\!\! } {\!\!\!\!\!\! } {\!\!\!\!\!\! }
\varphi^{\rm GNS}_{\nu_1 \nu_2| \nu_3 \nu_4}
\equiv
1 - (1 - \varphi^{(r)}_{\nu_1 \nu_2| \nu_3 \nu_4})
    (1 - \varphi^{(pp)}_{\nu_1 \nu_2| \nu_3 \nu_4})
\cr
\cr
&&
~~~
=
  \varphi^{(r)}_{\nu_1 \nu_2| \nu_3 \nu_4}
+ \varphi^{(pp)}_{\nu_1 \nu_2| \nu_3 \nu_4}
- \varphi^{(r)}_{\nu_1 \nu_2| \nu_3 \nu_4}
  \varphi^{(pp)}_{\nu_1 \nu_2| \nu_3 \nu_4},
\label{k19}
\end{eqnarray}
where the restriction parameters $\varphi^{(r)}$ and $\varphi^{(pp)}$
are those defined stochastically for rings,
Eq. (\ref{k10}), and for particle-particle
ladders, Eq. (\ref{k12}). Hence the reduced interaction
for this hybrid meets Kraichnan's conditions on label symmetry.

The effect of combining distinct classes of interaction in this
way is readily seen. When either class of phase factor survives,
its counterpart will not. If both combinations do survive
(as in the polarization to first and second order in the interaction)
there is no duplication. Their physics acts co-operatively
in the total correlation behavior, though never concurrently.

Diagrammatically, whether for the exact or any approximate
Hamiltonian, the functional $\Phi[\varphi V]$ for the
correlation energy is read off directly as the expectation
of the interaction part of the Hamiltonian. This involves the self-energy
$\Sigma[\varphi V; G]$, where $G[\varphi V]$ is the self-consistent one-body
propagator, or Green function. $\Phi$ and $\Sigma$ are related in two ways.
The first is via the Hellmann-Feynman integral identity
\cite{DP2}:
the underlying pair interaction $V$ is multiplied by a coupling
constant taken from zero to unity so
\begin{equation}
\Phi[\varphi V] \equiv {1\over 2}\int^1_0 {dz\over z} \sum_{\ell}
{\langle~ G_{-\ell}[z\varphi V] \Sigma_{\ell}[z\varphi V; G] ~\rangle} 
\label{k19.1}
\end{equation}
in which the self-energy and propagator within
the right-hand integrand are evaluated
at the coupling constant $z$. The second relation complementary
to Eq. (\ref{k19.1}) is the variational derivative
\cite{GB}
\begin{equation}
\Sigma_{\ell}[\varphi V; G]
= {\delta \Phi[\varphi V]\over \delta G_{-\ell}}.
\label{k19.2}
\end{equation}

The generic structure of $\Phi[\varphi V]$, whether exact
or associated with a Kraichnan Hamiltonian
or to its functional equivalent, $\Phi$-derivability
\cite{KB,GB}, has a very specific property.
Within the expansion of the exact $\Phi$ in
powers of the underlying potential partnered by the fully
renormalized propagators $G$ within the description,
each $G$ ``sees'' -- that is, is embedded in
-- a correlation environment identical to any other
propagator in the given term
\cite{GB}. It must not matter which $G$ is removed
to generate the self-energy diagrams for the relation Eq. (\ref{k19.2}).
The same $\Sigma$ must emerge.
Were the above not the case, $\Sigma$ would lack the symmetry needed
for conservation. Since its symmetry ultimately comes from the hermitian
nature of the Hamiltonian, it follows {\em a priori} and with no extra work
that every stochastic Hamiltonian model must possess a family of terms
making up $\Phi$ with the same symmetries as those that secure
microscopic conservation in the exact case.
This brings home the analytic power of Kraichnan's procedure.

\centerline{
\includegraphics[height=5truecm]{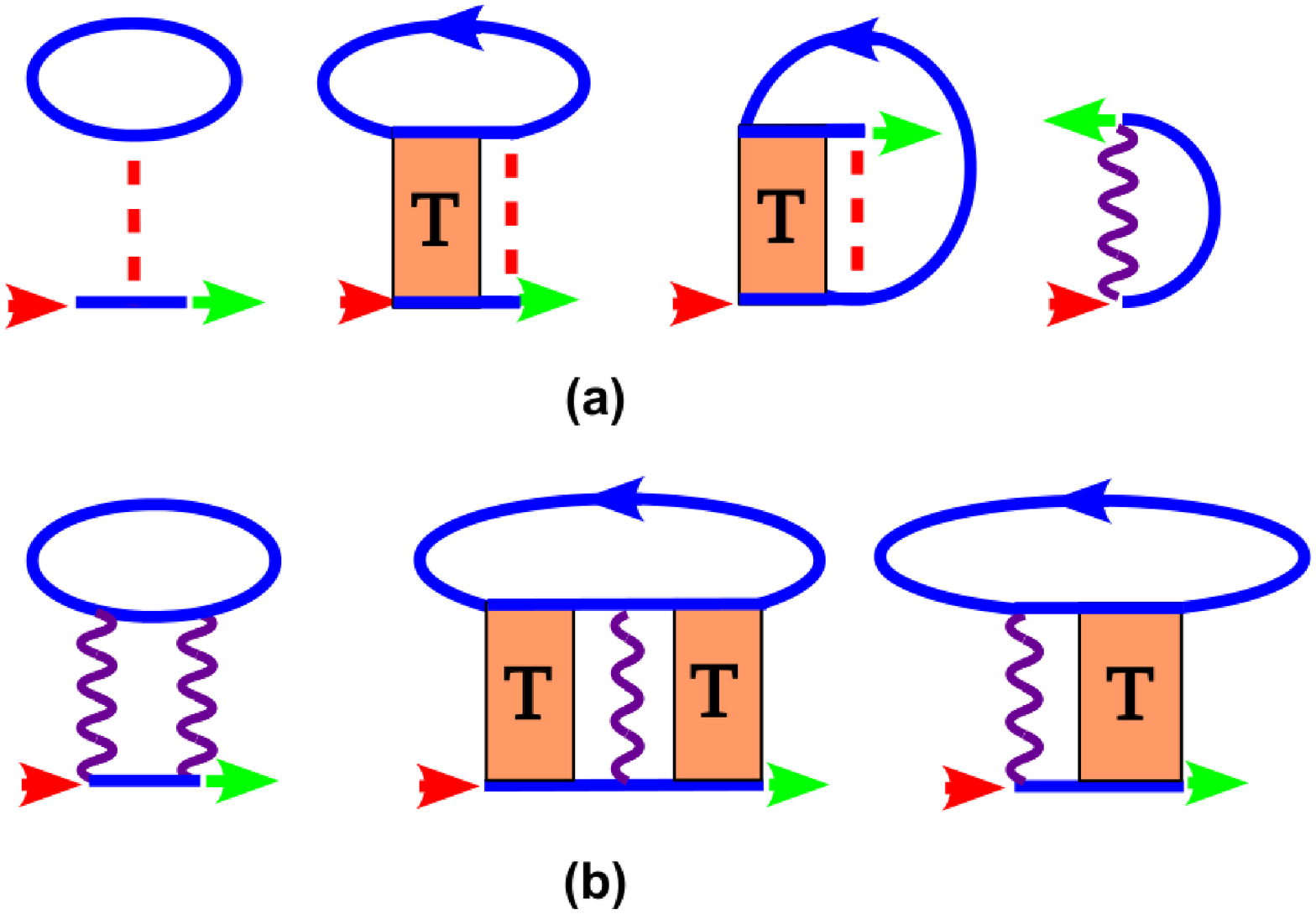}
}
{{\bf FIG. 8.} {\small Structure of model self-energy for the interpolating
rings-plus-ladders model. Allowed contributions appear in (a). Other
combinations, of which those in (b) are instances, are inhibited in the
stochastic average.}
}
\vskip 0.25cm

Figure 8 illustrates the consequences of the Hamiltonian
Eq. (\ref{k19}) for the self-energy given by Eq. (\ref{k19.2}).
Self-consistency of the one-body Green function through the self-energy
leads to the implicit nesting of rings and ladders
to all orders in the interaction. Nevertheless there can be
no ladders with chain-like rungs; they are stochastically suppressed.

The diagrams that survive stochastic filtering are
just those of the Green {\em et al}. prescription
\cite{GNS1}. Its polarization corrections to leading order in the
particle-particle $T$-matrix
are the sum of the terms in Figs. 4 and 6, compensated for overcounting.
Overcounting is automatically excluded in
Eq. (\ref{k19}) while in any constructive $\Phi$-derivable
model -- that of Ref. \cite{GNS1} is just one instance -- overcounting
must be corrected by hand because the choice of a correlation subset,
while obviously physically guided, is
still a matter of piece-by-piece selection out of the
full ground-state expansion.

Equation (\ref{k19}) furnishes the prototype for the similarly
motivated but more intricate approximations in the next Section.
With a proper Hamiltonian, treatment of various sum
rules in this model becomes much more efficient. The point is made
in Appendix A, in which the third frequency-moment sum rule is
recalled and interpreted in terms applicable to all models.

Having introduced the notion of selective combination of disparate
physical correlations within a unified Hamiltonian, we are ready
for more the comprehensive parquet and induced-interaction
series. Particularly in nuclear-matter and liquid-helium studies,
these distinguish topologically among particle-particle ladder processes,
sequential RPA-like polarization processes, and the latter's
exchange counterparts the particle-hole ladders. All three stochastic
components are available.

\section{Parquet and the Induced Interaction}

We end this paper with the discussion of Hamiltonians for approximate
theories based on a maximal inclusion of strictly pairwise correlations,
starting with the parquet theory. It has long been appreciated that not
all correlations in the many-body ground state are representable
as structures made up purely from sequential two-body scatterings.
Irreducible processes contribute that do not fit,
topologically, the templates covered above 
\cite{JLS}. Absent a general procedure to include these,
practical modeling efforts emphasized incorporating all possible
contributions reducible to the standard pair processes.

\subsection{Parquet Hamiltonian}

The most elaborate attempt at constructing a comprehensive theory
purely out of two-body processes is parquet.
A further significant feature of parquet is its intimate connection
with variational methods offering non-perturbative calculational
approaches to strong-correlation problems
\cite{JLS, JLS1}. The conceptual advantages of knowing its Hamiltonian
would go beyond the immediate precincts of diagrammatic theory.

The parquet diagrams include all those that, to all orders, would
tile the entire plane in systematic patterns; hence their name.
The ingredients for its Hamiltonian are at our disposal via the only
possibilities for two-body scattering: rings and the two species of
ladder, particle-particle and particle-antiparticle.
The parquet Hamiltonian is proposed to be
\begin{eqnarray}
&&
\!\!\!\!
{\cal H}^{\rm pqt}
\equiv
\sum_{\ell} \epsilon_k a^*_{\ell} a_{\ell}
+  {1\over 2N} \!{\sum_{\ell_1 \ell_2 \ell_3 \ell_4}}\!\!\!\!' ~~
\!\!\!
\varphi^{\rm pqt}_{\nu_1 \nu_2| \nu_3 \nu_4}
{\langle k_1 k_2 | V | k_3 k_4 \rangle}
a^*_{\ell_1} a^*_{\ell_2} a_{\ell_3} a_{\ell_4};
\cr
&&
\!\!\!\!
\varphi^{\rm pqt}_{\nu_1 \nu_2| \nu_3 \nu_4}
\equiv
1 - (1 - \varphi^{(r)}_{\nu_1 \nu_2| \nu_3 \nu_4})
(1 - \varphi^{(ph)}_{\nu_1 \nu_2| \nu_3 \nu_4})
(1 -  \varphi^{(pp)}_{\nu_1 \nu_2| \nu_3 \nu_4}).
\label{k21}
\end{eqnarray}
It is a generalization of the co-operative, yet strictly sequential,
structure of restriction factors in the rings-plus-ladders
Hamiltonian of the previous section, Eq. (\ref{k19}).
It manifestly allows for all possible planar topologies
produced by pairwise scatterings in maximally complex combinations
but not getting in one another's way, thereby ruling out any
diagrams that cannot be factorized in this sequential way.
As in ${\cal H}^{\rm GNS}$, overcounting cannot occur.

A formal demonstration that Eq. (\ref{k21}) yields the same
correlation structure as the standard formulation of parquet
is not pursued here.
What is already clear is that this proposal generates all self-consistent
admixtures of the three permissible scattering arrangements
for a many-particle system with a pair potential. The three core
processes operate sequentially, never concurrently, in any combination
generated from ${\cal H}^{\rm pqt}$.

For the reasons already noted for ${\cal H}^{\rm GNS}$
and illustrated in Fig. 8(b),
intermediate particle-hole processes are not permitted
within any particle-particle ladders for correlation diagrams derived
from ${\cal H}^{\rm pqt}$. One would need to check that this did not
restrict the parquet vertex structure
\cite{HB, JLS, JLS1, JLS2, JLS3} when interpreted, not as the diagrammatic
architecture directly seen in the ground-state correlation energy,
but indeed as its functional derivative
\cite{KB}; refer also to Eq. (\ref{A5})
of Appendix A. Confirmation that Eq. (\ref{k21}) leads to standard
parquet means reproducing the complete pair-scattering equations for
this variationally generated dynamical vertex,
to verify whether or not they are
identical to their parquet analogs.

\subsection{Induced Interaction}

The induced interaction
\cite{BB,AB} simplifies parquet by invoking
a parametrized effective pair potential to stand in for the
ladder sum of particle-particle scatterings. It has been effective
as a theory of static properties in hard-core Fermi systems
(nucleonic matter and noble-gas liquids) and their low-energy excitations
as well
\cite{AB}. A Coulomb-screened variant
has been applied to the low-density electron gas
\cite{AGP}.

In the induced interaction, explicit Brueckner-like
particle-particle scattering is omitted. Instead, the bare potential 
${\langle k_1 k_2 | V | k_3 k_4 \rangle}$ is replaced
with an antisymmetrized
approximation ${\langle k_1 k_2 | {\overline T}^{pp} | k_3 k_4 \rangle}$
to the ladders in Fig. 6(c); compare also Eq. (\ref{k07.1}). The Hamiltonian
includes only ring and particle-hole processes manifestly:
\begin{eqnarray}
&&
\!\!\!\!\!\!
{\cal H}^{\rm BB}
\equiv
\sum_{\ell} \epsilon_k a^*_{\ell} a_{\ell}
+ {1\over 2N} \!{\sum_{\ell_1 \ell_2 \ell_3 \ell_4}}\!\!\!\!' ~~
\!\!\!
\varphi^{\rm BB}_{\nu_1 \nu_2| \nu_3 \nu_4}
{\langle k_1 k_2 | {\overline T}^{pp} | k_3 k_4 \rangle}
a^*_{\ell_1} a^*_{\ell_2} a_{\ell_3} a_{\ell_4};
\cr
&&
\!\!\!\!\!\!
\varphi^{\rm BB}_{\nu_1 \nu_2| \nu_3 \nu_4}
\equiv
1 - (1 - \varphi^{(r)}_{\nu_1 \nu_2| \nu_3 \nu_4})
(1 - \varphi^{(ph)}_{\nu_1 \nu_2| \nu_3 \nu_4}).
\label{k22}
\end{eqnarray}

Figure 9 shows the essential ground-state correlation structure encoded in
Eq. (\ref{k22}). Now we construct a pair of dynamical
two-body scattering vertices,
$\Gamma$ for particle-hole and $\Xi$ for ring processes
(see Fig. 10),
following the induced-interaction template 
\cite{AGP,AB}.
Note that from now on a summation over an intermediate variable $\ell$
will be understood also to include intermediate integrals
in the frequency domain subject to conservation as for momenta.
In particular, the one-body causal propagator $G_{\ell}$ is now
in frequency-dependent form
\cite{rick}.

\centerline{
\includegraphics[height=6truecm]{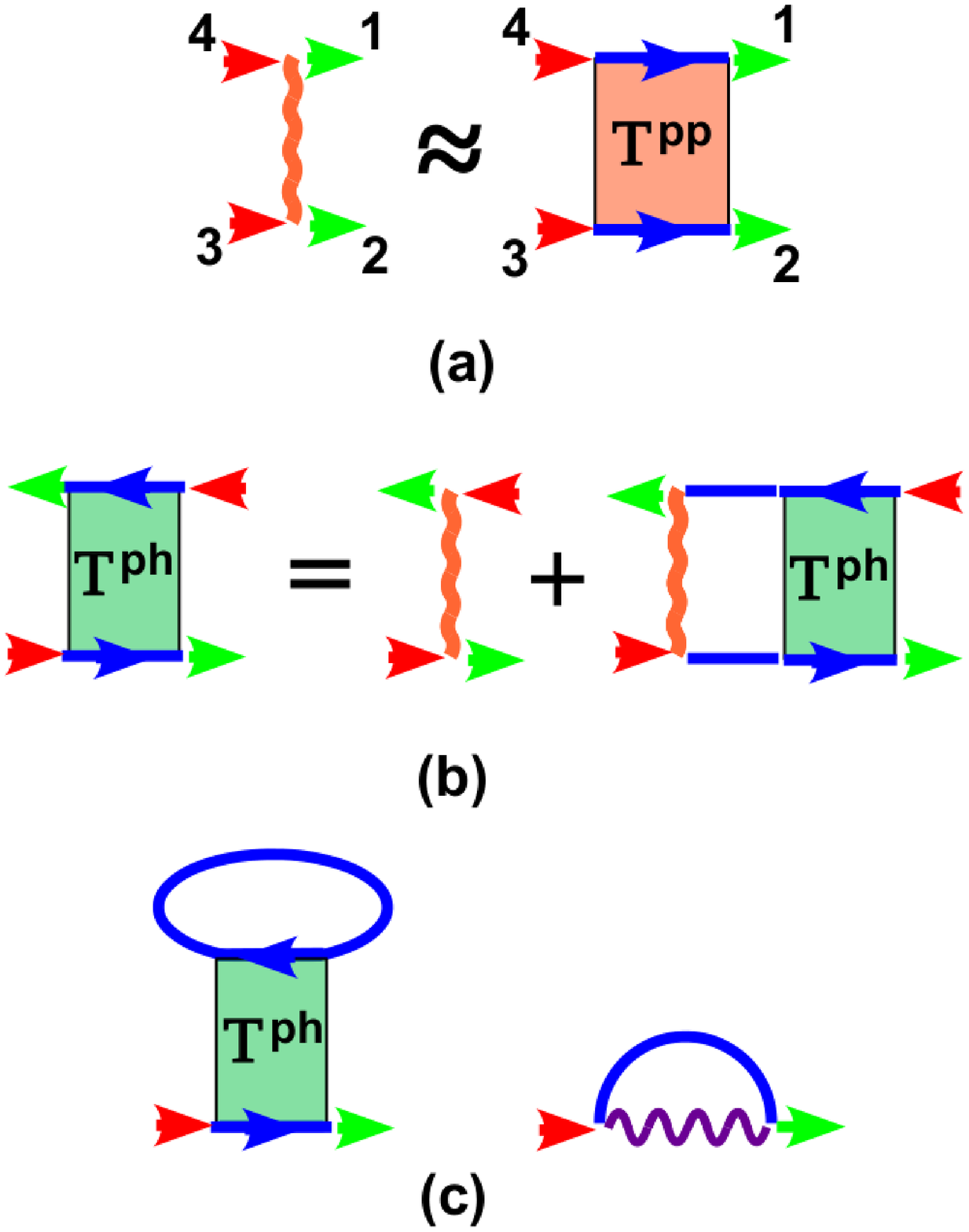}
}
{\bf FIG. 9.}
{\small Definition of the induced-interaction approximation.
(a) The particle-particle $T$-matrix, or ladder vertex, is replaced with an
antisymmetrized
effective potential ${\overline T}^{pp}$.
(b) The particle-hole ladder series is defined by its
Bethe-Salpeter equation; intermediate one-body propagators are
self-consistently defined within the approximation as a whole.
(c) The self-energy derived from the interaction Hamiltonian is
determined by the particle-hole $T$-matrix, selected via $\varphi^{(ph)}$,
while the shielded interaction is selected through
$\varphi^{(r)}$; the latter
is defined as in Fig. 4(c) except that the bare potential $V$ is
replaced with the particle-particle Ansatz
${\overline T}^{pp}$ schematized in (a) above. 
}

The effective vertices $\Gamma$ and $\Xi$ subsume
all non-canceling internal responses to an external disturbance.
They include, but are not the same as, the vertex
appearing in the equilibrium self-energy whose induced-interaction
form is shown in Fig. 9(c).
Rather, they correspond to derived two-body scattering processes
implicit in the correlation energy functional $\Phi$ but made
manifest only through the dynamic response of the system
\cite{KB}.
Appendix A details the behavioral difference between the
differently structured vertices.

The derived dynamical vertices should sum, consistently,
all those intermediate two-body scatterings assured of
surviving the internal stochastic averaging over
$\varphi^{(ph)}$ and $\varphi^{(r)}$.
Thus the $\Gamma$ candidate is defined to have the structure
\begin{eqnarray}
&&
\!\!\!\!\!\! \!\!\!\!\!\! \!\!\!\!\!\!
\!\!\!\!\!\! \!\!\!\!\!\! \!\!\!\!\!\!
\Gamma{(\ell_1 \ell_2 | \ell_3 \ell_4)}
\equiv
{\langle k_1 k_2 | {\overline T}^{pp} | k_3 k_4 \rangle}
+ (\varphi^{(ph)}_{\nu_1 \nu_2| \nu_3 \nu_4})^{-1}
\!{\sum_{\ell'_1 \ell'_2 \ell'_3 \ell'_4}}\!\!\!\!' ~~
\Gamma{(\ell_2 \ell'_1| \ell_4 \ell'_3)}
%
\cr
\cr
&&
~~~ ~~~ ~~~ ~~~ 
\times
\varphi^{(ph)}_{\nu_2 \nu'_1| \nu_4 \nu'_3}
(-\delta_{\ell'_1 \ell'_4} \delta_{\ell'_2 \ell'_3}
G_{\ell'_1} G_{\ell'_2})
\varphi^{(ph)}_{\nu_1 \nu'_2| \nu_3 \nu'_4}~
\Xi{(\ell_1 \ell'_2| \ell_3 \ell'_4)}
%
\cr
\cr
\cr
&&
\!\!\!\!\!\! \!\!\!\!\!\! \!\!\!\!\!\! \!\!\!\!\!\!
=
{\langle k_1 k_2 | {\overline T}^{pp} | k_3 k_4 \rangle}
-{\sum_{\ell}}
~\Gamma{(\ell_2 \ell| \ell_4 \ell')}
G_{\ell} G_{\ell'}
\Xi{(\ell_1 \ell'| \ell_3 \ell)};
~~\ell' = \ell + \ell_3 - \ell_1.
\label{k25}
\end{eqnarray}
%
This corresponds to the sum of ``$t$-channel irreducible''
processes
\cite{AB}, namely those that cannot be separated into two sub-vertices
by cutting any particle-hole line pair with momentum transfer
$k_1 - k_3$.
The negative sign in the summation on the right-hand side
is due to exchange of one pair of particle (or hole) labels,
relative to the complementary ring-like vertex $\Xi$; see
Eq. (\ref{k26}) below. Any  stochastic average with $\varphi^{(r)}$,
for the object $\varphi^{(ph)}\Gamma$,
will be suppressed owing to the vertex topology. 

A concomitant summation gathers all ring-like
scatterings defining $\Xi$, so
\begin{eqnarray}
&&
\!\!\!\!\!\! \!\!\!\!\!\! \!\!\!\!\!\!
\!\!\!\!\!\! \!\!\!\!\!\! \!\!\!\!\!\!
\Xi{(\ell_1 \ell_2| \ell_3 \ell_4)}
\equiv
\Gamma{(\ell_1 \ell_2| \ell_3 \ell_4)}
+ (\varphi^{(r)}_{\nu_1\nu_2| \nu_3 \nu_4})^{-1}
\!{\sum_{\ell'_1 \ell'_2 \ell'_3 \ell'_4}}\!\!\!\!' ~~
\Gamma{(\ell_1 \ell'_2 | \ell'_3 \ell_4)}
%
\cr
\cr
&&
~~~ ~~~ ~~~ ~~~ 
\times
\varphi^{(r)}_{\nu_1 \nu'_2| \nu'_3 \nu_4}
(\delta_{\ell'_1 \ell'_3} \delta_{\ell'_2 \ell'_4}
G_{\ell'_1} G_{\ell'_2})
\varphi^{(r)}_{\nu'_1 \nu_2| \nu_3 \nu'_4}~
\Xi{(\ell'_1 \ell_2 | \ell_3 \ell'_4)}
\cr
\cr
\cr
&&
\!\!\!\!\!\! \!\!\!\!\!\! \!\!\!\!\!\! \!\!\!\!\!\!
=
 \Gamma{(\ell_1 \ell_2 | \ell_3 \ell_4)}
+ {\sum_{\ell}}
~\Gamma{(\ell_1\ell'' | \ell \ell_4)}
G_{\ell} G_{\ell''}
\Xi{(\ell \ell_2 | \ell_3 \ell'')};
~~\ell'' = \ell + \ell_3 - \ell_2.
\label{k26}
\end{eqnarray}
If we attempt an operation involving a stochastic
average over $\varphi^{(ph)}$ of the RPA-like object
$\varphi^{(r)}(\Xi - \Gamma)$, corresponding to the induced interaction's
``$u$-channel irreducible'' series (not separable
into two sub-vertices by cutting any particle-hole line pair
with momentum transfer $k_2 - k_3$), the result will be suppressed.
Inspection of the series expansion of the latter shows
that Eq. (\ref{k26}) has the symmetry
\[
\Xi{(\ell_1\ell_2| \ell_3 \ell_4)}
= 
 \Gamma{(\ell_1\ell_2| \ell_3 \ell_4)}
+ {\sum_{\ell}}
~\Xi{(\ell_1\ell'' | \ell \ell_4)}
G_{\ell} G_{\ell''} \Gamma{(\ell \ell_2| \ell_3 \ell'')}.
~~~
\]
Furthermore, substituting $\Gamma$ from
Eq. (\ref{k25}) into the right-hand side
of Eq. (\ref{k26}) renders $\Xi$ explicitly
antisymmetric under pair exchange
\cite{AB}.
The coupled structure of Eqs. (\ref{k25}) and (\ref{k26}) is shown
in Fig. 10.
\vskip 0.15cm
\centerline{
\includegraphics[height=4truecm]{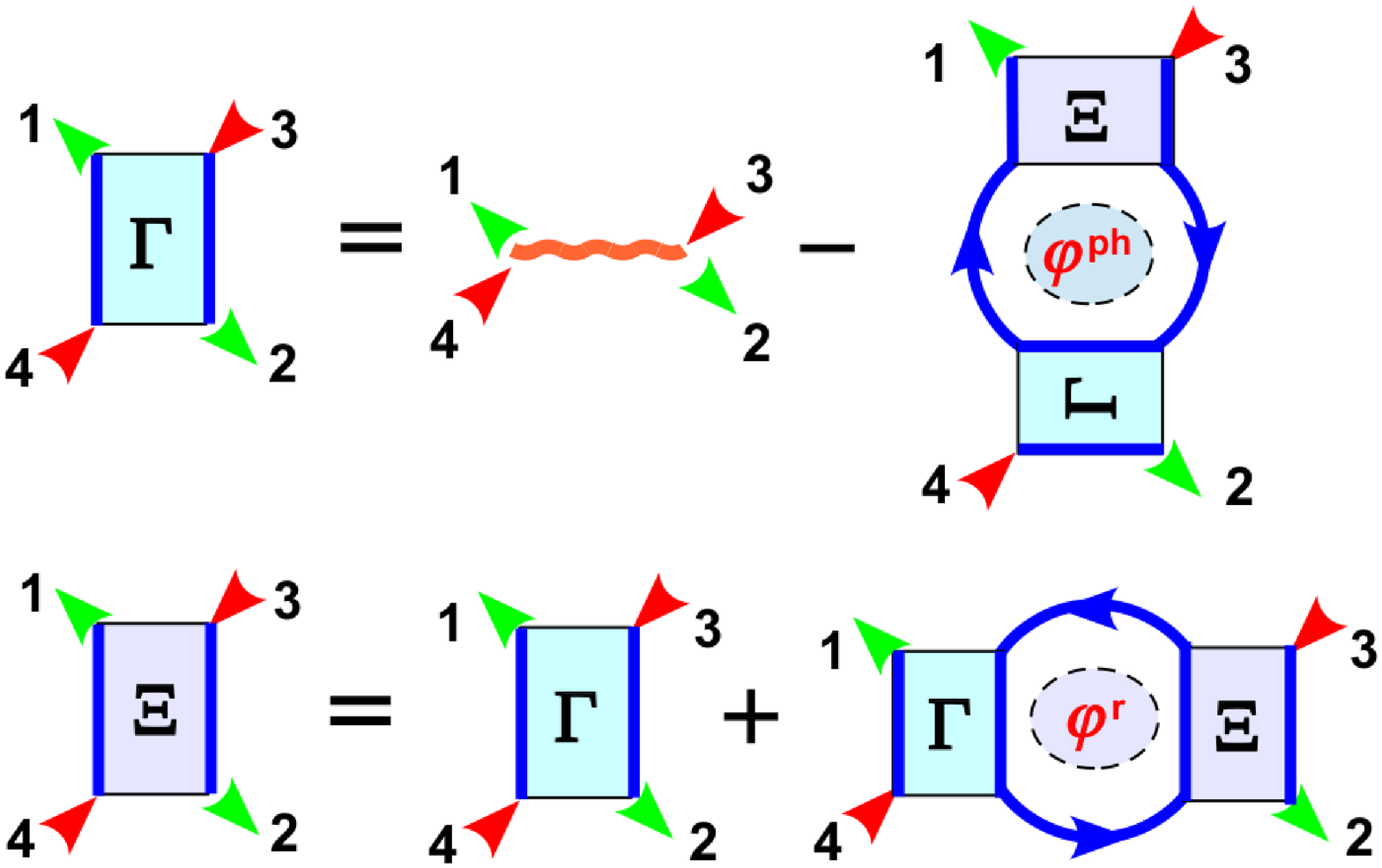}
}
{{\bf FIG. 10.} {\small Particle-hole- and ring-like vertices
mediate the dynamical interaction between particle and hole pairs in the
induced-interaction model after its Hamiltonian, Eq. (\ref{k22}).
These processes determine the system's self-consistent response
to an external perturbation.
The topology of the two-body vertex $\Gamma$ sums all intermediate processes
that are not automatically suppressed by stochastic averaging of its
accompanying restriction factor $\varphi^{(ph)}$. Correspondingly,
the interaction vertex $\Xi$ includes all possible topologies
that are not automatically suppressed by an average over the rings-only
factor $\varphi^{(r)}$. Also note that the phenomenological particle-particle
vertex ${\overline T}^{pp}$ is antisymmetrized for particle-pair exchange
$1 \leftrightarrow 2$ or $3 \leftrightarrow 4$. Thus the
complete induced-interaction scattering amplitude
$\Xi$ is itself antisymmetric.
}}
\vskip 0.25cm
After averaging independently over the two stochastic
restriction factors, the vertex $\Xi$ emerging from
Eqs. (\ref{k25}) and (\ref{k26}) leads
to the set of dynamical two-body scattering processes
within the induced-interaction model.
As they stand, prior to any stochastic averaging, our vertex equations
neglect all terms carrying the restriction factors $\varphi^{(r)}$
and $\varphi^{(ph)}$ concurrently.
\centerline{
\includegraphics[height=3.5truecm]{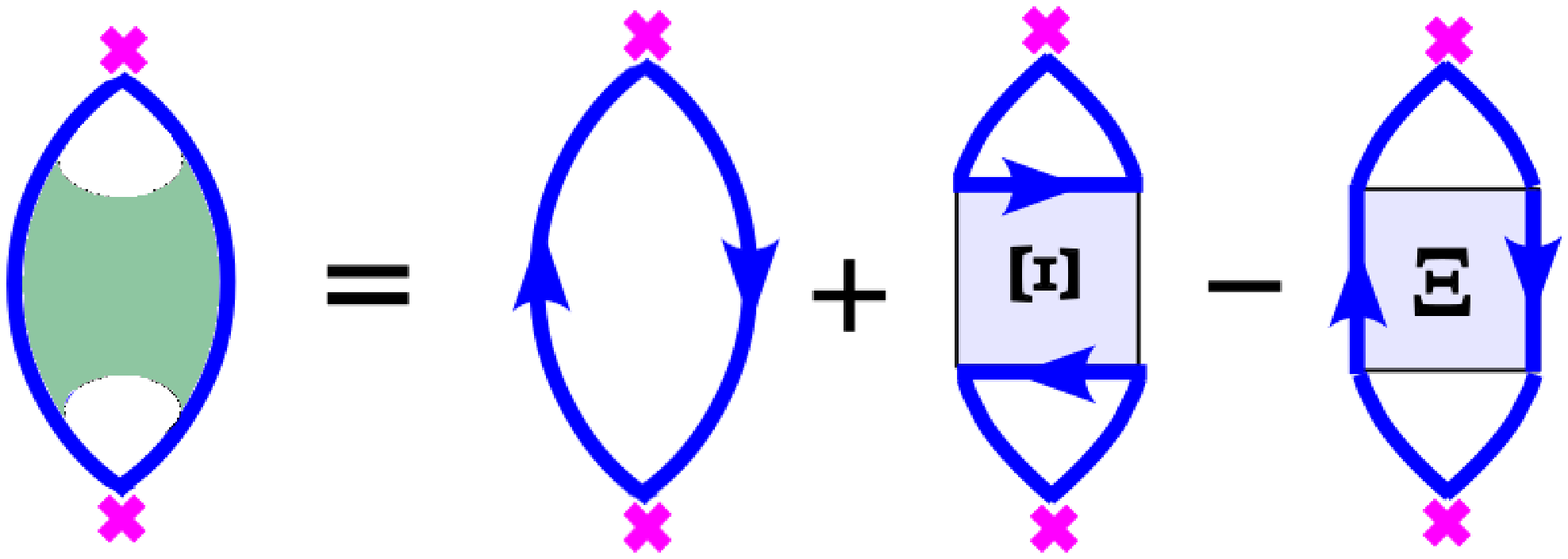}
}
{{\bf FIG. 11.} {\small Total density-density response function
determined by the self-consistent two-body vertex structure in
the induced-interaction model. Crosses indicate coupling to a weak
external perturbing potential. (If coupling is to the current,
its operator attaches to the external vertices and
the diagram describes the current autocorrelation function.)
The leading right-hand term is the renormalized polarization
with no particle-hole vertex; the second incorporates
the contributions generalizing Hartree-like (RPA) screening;
the last term holds the complementary particle-hole
ladders responsible for Fock-like exchange scattering.
}}
\vskip 0.25cm
\noindent
These remain legitimate parts of the
complete $\Xi$, until the final average; when this
is performed, the terms expressly left out of the coupled
self-consistent pair Eqs. (\ref{k25}) and (\ref{k26}) are precisely
those that vanish by destructive interference.
Then $\Xi$ becomes
the induced-interaction vertex bearing the dynamic correlations
in the model and determining its response functions, such as
$\chi({\bf q},\omega)$, exhibited in Fig. 11.

\section{Summary}

The goal of this paper has been the rational construction of
explicit Hamiltonians for significant conserving approximations
lacking them, in problems of strongly interacting assemblies. Chief among
the many-body problems of interest are short-range dynamics in
charged quantum fluids such as the electron gas, as well as
nuclear matter and the noble-gas fluids.

Conventionally, diagrammatic theories of correlations have been
set up via other microscopic prescriptions, such as
$\Phi$-derivability; but methods that build their correlation structure
heuristically from the bottom, so to speak, do not generate a
Hamiltonian corresponding to their model.
This can make it problematic to confirm essential canonical properties,
notably the conserving sum rules, which are hallmarks of the exact
theory and which one wants to validate equally for any approximate
description.

A systematic strategy for constructing model Hamiltonians was formulated
by Kraichnan. It consists in (\i) embedding the exact interacting
problem within a large ensemble of identical but distinguishable
system copies, (\i\i) adjoining, to their exact interaction potential,
randomly chosen factors coupling stochastically all the copies
in the collection, and (\i\i\i) designing the coupling scheme
so that only specific, restricted sets of expectation values for
correlations will survive stochastic
averaging over the introduced couplings. All other combinations
will be suppressed by destructive interference and vanish in the limit
of an infinite ensemble average.

Correlations selected in that way will bring out
the effects believed to prevail in a given physical context.
For example, one form of stochastic coupling will pick out screening
correlations in a characteristically long-ranged Coulomb system. Another
form will promote repeated particle-particle scattering in systems
with a hard-core potential.

First, the technicalities of Kraichnan's construction were recalled.
Next came a survey of applications originally given by Kraichnan. Included
were the random-phase and Hartree-Fock approximations and their refinement
in the shielded potential, or ring, model, and the ladder series for
hard-core systems such as nuclear matter. These steps set the scene for
the third part: adaptation of the stochastic method to more elaborate
correlation theories for which a Hamiltonian has not been at hand.

Three approximations of interest were discussed and Hamiltonians were
identified for them. All involve a microscopically consistent unification
of short-range with long-range correlations.
They are: the ring-plus-ladder model, the parquet theory,
and the induced-interaction construction. For the latter an explicit
pair of particle-hole dynamical scattering-vertex equations was described,
based on a generalized definition of the stochastic coupling factor.
I showed that the vertex equations definable within the stochastic Hamiltonian
formalism are the same as their heuristic counterparts establishing the
induced interaction.

Appendix A examines the role and interpretation of the sum rules in
conserving models, concentrating on the third frequency-moment sum rule.
The sum-rule structure for a conserving approximation follows canonically
from its Hamiltonian (when known) inheriting its analytic properties from
the complete system description. However, care has to be taken with how
these relations are evaluated and interpreted. Other identities that are
not sum rules and valid for the exact system, need not hold in an
approximation
\cite{JS}; the price of any simplification.
Still, an advantage of knowing the Hamiltonian is
automatic validity for all sum rules that come
out of microscopic conservation plus the causal boundary conditions.
One has only to apply the rules discerningly.
Appendix B contains brief remarks on possible relations between the
existence of approximate interacting Hamiltonians and complementary
nondiagrammatic solutions to general correlation problems.

Future work would include a demonstration that the proposed
parquet Hamiltonian
Eq. (\ref{k21}) yields a dynamical two-body vertex structure
identical to that originally worked out in the parquet literature.
To the extent that parquet in particular has an intimate link
to nonperturbative variational methods in strong correlations
\cite{JLS,JLS1}, any consequences of confirming
the parquet Hamiltonian follow through for those approaches.
At a more general level, as sketched in Appendix B,
similar considerations might be applied to any interacting model
reliant on an underlying Hamiltonian.
One could also explore how Kraichnan's stochastic
Hamiltonians may apply with increasing sophistication and physical
fidelity beyond linear response and in lower dimensions
\cite{RHK2, wims},
not only in uniform Coulomb systems but in inhomogeneous interacting
systems of all types. As a conceptual tool, some of its power may have
been demonstrated in this paper. As a practical tool it awaits further
thought.

\section*{Acknowledgments}

I thank Professor Kenneth Golden for stimulating my return to this
long-standing problem, Professor Mukunda Das for his forthright
and invariably fruitful comments as the work developed, and
Professor Alexander Lande for directing me to Reference
\cite{JLS3} and its extension of the parquet approach to three-body
correlations.

\appendix
\section{Third frequency-moment sum rule}

The importance of the third frequency-moment sum rule for
short-range correlation properties, Coulomb fluids included, 
was first highlighted by Goodman and Sj\o lander
\cite{GS}.
They gave a proof of the rule and analyzed the information
it contains about the near environment, or ``correlation hole'',
of a typical particle within its interacting medium.

Here we focus upon the relevance of this sum rule as a paradigm
for the way in which approximate correlation models, despite
being assured of satisfying the sum rules of the full case, call for a more
careful understanding of what the sum rules may have to tell.
We will not detail the proof of the third-moment rule, relying
on Ref.
\cite{GS}; a more diagrammatically oriented proof is in Ref.
\cite{GNS3}.

We state the rule as it applies to the electron fluid.
If one takes the dynamic and static structure factors for the system
\cite{DP2,GS}, respectively  $S(q,\omega)$ and $S(q)$,
the third-moment rule is
\begin{eqnarray}
&&\int^{\infty}_{-\infty} \!{d\omega\over 2\pi}
\omega^3 S(q, \omega)
=
{q^4n\over 2} M_3(q);
\cr
\cr
&&
\!\!\!\!\!\! \!\!\!\!\!\! \!\!\!\!\!\! \!\!\!\!\!\!
\!\!\!\!\!\! \!\!\!\!\!\! \!\!\!\!\!\! 
M_3(q)
\equiv
{q^2\over 4} + 2\sum_k \!\varepsilon_k {\langle a^*_k a_k \rangle}
\!+\! V(q)n {\Bigl[ 1 \!-\! n^{-1} \sum_{q'} ({\bf\hat q}\cdot{\bf\hat q'})^2
[ S(q') \!-\! S(|{\bf q} \!-\! {\bf q'}|) ]  \Bigr]}.
\label{A1}
\end{eqnarray}
Unlike the first-moment ({\em f}-sum) rule, this
identity gives weight to the high-frequency (short-time,
thus also short-ranged) properties of the assembly.
Hence it is much more sensitive to the correlation
structure. That is evident through the
second right-hand term of the factor $M_3(q)$, which is the expectation
of the kinetic energy over the interacting Fermi sea.
Sensitivity to correlations comes out even more clearly
in the last contribution, explicitly dependent on the
static structure factor whose nature we now discuss.

In addressing the structure factor $S(q)$ we make an important observation.
Through its manifest sensitivity to the correlations in the system,
Eq. (\ref{A1}) for the third-moment rule will equally reflect the
correlation properties of any approximation to the exact physics.
By that it may also accentuate the physical shortcomings
of the approximation, so the rule is an important quantitative gauge
of a model. The latter does not touch
the architecture of the rule, which remains valid; it means
that one must be careful how the right- and left-hand sides of
Eq. (\ref{A1}) have to be evaluated. 

Commonly termed the ``static'' structure factor, $S(q)$ is the
instantaneous pair correlation function in Fourier space:
\begin{equation}
S(q) \equiv \sum_{kk'} {\langle a^*_{k+q}a_k a^*_{k'-q} a_{k'} \rangle}.
\label{A2}
\end{equation}
Mathematically it is generated by direct removal of an interaction
line in the diagrammatic expansion for $\Phi$:
\begin{equation}
S(q) \equiv {\delta \Phi\over \delta V(-q)};
\label{A3}
\end{equation}
its inner structure therefore represents the equilibrium
correlation structure directly
\cite{GNS3}. By contrast, the dynamic structure factor is the density
response to a weak, but external, perturbation itself coupling to the
density. This is in sharp functional distinction to $S(q)$, which is
strictly determined in the ground state.
We stress that $S(q,\omega)$ is not an equilibrium property
although it is computed in terms of equilibrium expectation values.

Now we look at how $\Phi$ is perturbed. 
A weak external potential $U$ couples to the density operator
through a one-body term $U(q,\omega)a^*_{k+q}a_k$ added to the Hamiltonian,
Eq. (\ref{k01}). This changes the correlation energy:
\begin{eqnarray}
\Phi[U]
=
\Phi[0] \!+\! {1\over 2} U^*(q,\omega) \chi(q,\omega) U(q,\omega)
\!+\! {\cal O}(|U|^4);
~~~~~
\label{A4}
\end{eqnarray}
there is no linear term since $\Phi$ is a minimum at equilibrium.
However, obtaining the response function is no longer a simple matter
of removing an interaction line from $\Phi$, as for $S(q)$.
We must track down every occurrence of $U$ including its appearance
in the self-consistently recurrent structure of the propagators $G[U]$.
A clear and very detailed exposition of the process is in Refs.
\cite{KB} and
\cite{GB}.

Let $\Lambda[G]$ be the vertex defining the correlation energy
so that, symbolically, the perturbed self-energy is
$\Sigma \equiv U + \Lambda[G]\!:\!G[U]$ where for brevity
we denote by ``$:$''
internal integrations over momentum-energy. The dynamic response
to lowest order in the perturbation is encoded in the quantity
\begin{eqnarray}
&&
\!\!\!\!\!\! \!\!\!\!\!\! \!\!\!\!\!\! \!\!\!\!\!\! \!\!\!\!\!\!
\delta \Phi
=
{1\over 2} U^*:
{\left[ {\delta G\over \delta U}
+ {\delta G\over \delta U^*}: 
{\delta^2\Phi\over \delta G\delta G'}
:{\delta G'\over \delta U} \right]}:U
\cr
\cr
&&
\!\!\!\!\!\! \!\!\!\!\!\! \!\!\!\!\!\! \!\!\!\!\!\!
\!\!\!\!\!\! \!\!\!\!\!\! \!\!\!\!\!\!
=
{1\over 2} U^*\!\!:\!{\left[ {\delta G\over \delta U}
\!+\! {\delta G\over \delta U^*}\!:\!\Lambda\!:\!{\delta G'\over \delta U}
\!+\! {\left( {\delta G\over \delta U^*}\!:\!
{\delta \Lambda\over \delta G''}\!:\!{\delta G''\over \delta U}:G'
\!+\! G\!:\!{\delta G''\over \delta U^*}\!:\!
{\delta \Lambda\over \delta G''}\!:\!
{\delta G'\over \delta U}
\right)} \right]}\!\!:\!U.
\label{A5}
\end{eqnarray}
Aside from the leading term ${\delta G/\delta U} = GG$
on the right-hand side of the
second expression (the renormalized zeroth-order polarization),
comparison of Eqs. (\ref{A4}) and (\ref{A5}) shows that
the diagrammatic structure of the dynamic response
$\chi(q,\omega)$ is not solely determined by that of $\Lambda$,
whose terms appear in the ground-state energy functional
directly defining the conserving one-body $\Sigma[G]$,
but also, and crucially
for microscopic conservation at the two-body level, by the
new contributions generated through self-consistency
of the correlations in the system
\cite{KB,dubois}. The phenomenon
is illustrated in Figs. 4(b), 6(b), and 7(b) for the three primary
models of Sec. III and in Fig. 11 of Sec. V.

The central message of this discussion is that,
for any description of a correlated
system, $S(q,\omega)$ as the negative imaginary part of
$\chi(q,\omega)$
has explicit extra terms appearing in it that are otherwise
dormant in the ground-state energy functional. In any approximate
picture of correlations, in other words, the dynamical vertex
and its $S(q,\omega)$ on the one hand will not have the same
diagrammatic structure as the ground state and its $S(q)$ on the other.

These objects lead to quite different results.
This does not contradict the fact that all the sum rules
that apply to the full theory -- including the correlation-sensitive
third-moment rule -- remain valid in any approximation
built on the Kraichnan or functionally equivalent $\Phi$-derivable
pattern. It comes down to a consistent reading of the sum rules.

Equation (\ref{A1}) in any approximate model is interpreted
correctly if, and only if, the dynamic structure factor on the
left-hand side derives from Eq. (\ref{A5}) while, on the
right-hand side, the static structure factor is obtained from
Eq. (\ref{A3}). That is because, in diagrammatic terms,
the prime physical basis of $S(q)$ resides directly in the
ground-state properties through $\Lambda$
\cite{GNS3}; true in the exact case, thus true
for any properly constituted approximation.

Confusion has sometimes arisen over this conceptual point, not just for
the third-moment sum rule but for other instances such as
the compressibility sum rule
\cite{VS}. In the exact theory --
and in the exact theory alone -- the
static factor $S(q)$ has another, possibly more
familiar, expression as the frequency integral 
of $S(q,\omega)$
\cite{DP2}:
\begin{equation}  
S(q) = \int^{\infty}_0 d\omega S(q,\omega).
\label{A6}
\end{equation}
In experiment this relation gives the scattering cross-section
from an angle-resolved measurement uncollimated for
inelastic energy loss $\omega$, whose cross-section as measured
would be $S(q,\omega)$. 
Interpreted theoretically, its strongly model-dependent form is not
a sum-rule identity obtained from standard arguments
using analyticity and the Kramers-Kr\"onig relations
\cite{mahan}, whose causal structure is immune
to ensemble averaging.

If applied in any {\em approximation} to the full problem,
Eq. (\ref{A6}) fails to yield the same result as Eq. (\ref{A3}).
For the RPA, Eq. (\ref{A3}) results in a trivial pair correlation
function in real space with no features at all, while Eq. (\ref{A6})
for RPA results in a pair correlation function
that becomes unphysically negative
\cite{mahan}.
Thus, by itself, formal conservation hardly secures good
numbers in a model; but
feeding the evaluation of Eq. (\ref{A6}) into the right-hand
side of Eq. (\ref{A1}) makes matters worse by breaking
sum-rule consistency.

Suppose we had obtained, from
Eq. (\ref{A3}), a poor estimate for $S(q)$ compared to measurement.
We might turn to Eq. (\ref{A6}), somewhat unsystematically in this context,
expecting a better answer (with no guarantee of
improvement). Unfortunately this forfeits its canonical pedigree from
the model Hamiltonian because the third-moment sum rule
would be violated with that choice. 

As far as is known the equivalence of (\ref{A3}) and (\ref{A6})
is only for the exact ground state
\cite{JS}. The reason appears to be the dependence of
Eq. (\ref{A6}) on the Fermi golden rule
\cite{DP2}, itself exploiting completeness of
the many-body eigenstates in Fock space.
In the Kraichnan ensemble average, the contribution of
whole families of states is washed out (albeit, prior to averaging,
the completeness of Fock space holds for each individual
member in the ensemble of stochastic Hamiltonians).
This kills the state coherence essential to Eq. (\ref{A6}).

It is reasonable to surmise
that the distinction between a set of virtual (dormant) dynamic
correlations in $S(q)$ and their real manifestation in
$S(q,\omega)$, mandated by conservation
\cite{KB}, applies to the actual exact description.
Then Eq. (\ref{A6}) reveals a deeper and extremely rigid constraint
on the terms beyond those in $\Lambda$ on the right-hand
side of Eq. (\ref{A5}), impossible to meet within any approximation
\cite{JS}.
The discrepancy between the two evaluations of $S(q)$, canonical
for Eq. (\ref{A3}) but in practice empirical for (\ref{A6}), is
the price paid by any truncation of the full problem,
no matter how elaborate.
Indeed it could be used as an inbuilt measure
of the mismatch between a reduced correlation theory and its
fully correlated parent.

\section{Kraichnan's construction and nondiagrammatic analyses}

This addendum remarks informally on how the
stochastic-Hamiltonian approach may relate to
self-consistent, correlated theories
not reliant on diagrammatic analysis
characteristic of Green-function methodology.
We focus on two nonperturbative examples: density-functional theory
\cite{dft1,dft2} and the coupled-cluster formalism
\cite{cc1,cc2}.

\subsection{Density-functional theory}

Basic to density-functional theory (DFT)
is the proof that the ground-state energy expectation of a
many-body system interacting in the normal state is a unique
functional of its particle-density distribution
\cite{dft1}.
Then, given an independent constitutive relation between particle density
and exchange-correlation energy density, the problem of determining
the interacting system's behavior can be closed and solved.

There are many ways to negotiate approximate closures for DFT;
but the {\em exact} formulation of its basic Hohenberg-Kohn
and Kohn-Sham theorems
\cite{dft1} is not negotiable. The question arises whether there
exist physically meaningful approximations to correlated systems
for which the foundational DFT theorems are equally valid.
That indeed there are such models was established by Langreth
\cite{dft2}.

In brief, Langreth demonstrates that any $\Phi$-derivable correlation
model (its exchange-correlation energy functional
meets the Baym-Kadanoff criteria for microscopic conservation
\cite{KB,GB}) will satisfy the DFT theorems. Hence any
method for solving the DFT equations is applicable to this
wide class of model. These offer a different quality and
order of approximation over and above strategies such as local-density
and generalized-gradient methods
\cite{dft1}, ordinarily invoked to solve density-functional problems.

The present paper has shown how extended models of
correlations based on Kraichnan's Hamiltonian structures
are equivalent to the $\Phi$-derivable description of their
free-energy functional. From Ref. 
\cite{dft2} it follows that there are physically
nontrivial density-functional theories that, while approximate,
possess a fully defined and valid Hamiltonian in the sense of Kraichnan.
Any useful implications for DFT praxis fall outside our ambit here
and would need closer study within that specific context.

\subsection{Coupled-cluster method}

The situation of the coupled-cluster method (CCM)
(also known as ``exp$S$'')
\cite{cc1,cc2}, vis \`a vis
the existence of model Hamiltonians,
at first glance is not dissimilar to the case of DFT.
One way to make a connection is to
note that the coupled equations defining the exp$S$ method 
\cite{cc1}
address certain overlap integrals for the Hamiltonian,
selecting the set of ``linked'' amplitudes that determine
the irreducible contributions to the correlation-energy
functional $\Phi[V]$, already discussed in Sec. IV.

If one embeds the Hamiltonian of CCM using
the Kraichnan prescription for its interaction part Eq. (\ref{k06}),
along with a physically guided choice for the restriction parameters
$\varphi_{\nu_1 \nu_2 | \nu_3 \nu_4}$, then
an appropriate coupled-cluster formulation exists
{\em for each member of the Kraichnan ensemble} as well as collectively.
Subsequently this assembly would be subject to stochastic averaging
just as in the diagrammatic approach. One would need to ask
how the exp$S$ equations changed in any process of reduction
and, importantly, whether the operation of taking overlap
integrals in exp$S$ should be expected to commute with that
of stochastic averaging; for, the
same distinction seen in Appendix A would arise between
canonical procedures immune to stochastics and those
sensitive to the accompanying loss of state completeness.

Answers to these issues may lie in the fact that
the (nondiagrammatic) coupled-cluster formalism has a correspondence to the
(diagrammatic) Goldstone time-ordered expansion
\cite{cc1} and therefore in principle has a path back to $\Phi$-derivability
\cite{GB}. Further investigations along such lines would be enlightening.

A different aspect of CCM is its hierarchical truncation of linked amplitudes
(``SUB2'', ``SUB3'' etc. within the terminology). It is difficult
to tell -- at this point -- whether suitable Kraichnan restriction
parameters $\varphi$ could be systematically defined for these.
A possible analogy is the treatment of
RPA and Hartree-Fock within the Kraichnan approach (see Secs. III A and B)
where truncation via their corresponding $\varphi$
is not stochastic at all, but simply sets to zero
anything beyond those two basic correlations. Even diagrammatically 
it is an open question whether something equally prescriptive
and not stochastic operates at higher orders of correlation.
It suggests a very interesting problem that could also, in its turn,
shed light on the nature of Kraichnan's program itself.

\section*{References}

\end{document}